\documentclass{aa}
\usepackage{natbib,lineno} \bibpunct{(}{)}{;}{a}{}{,} 
\usepackage{graphicx}
\usepackage{color}

\begin{document}

\title{Making Faranoff-Riley I radio sources.} 
\subtitle{I. Numerical hydrodynamic 3D simulations of low-power jets}

\author{S. Massaglia\inst{1}, G. Bodo\inst{2}, P. Rossi\inst{2}, S. Capetti\inst{2} \and  A. Mignone\inst{1} }

\authorrunning{S. Massaglia et al.} 
\titlerunning{Making Faranoff-Riley I radio sources}

\institute{ Dipartimento di Fisica, Universit\`a degli Studi di Torino, via Pietro Giuria 1, 10125 Torino, Italy \and INAF/Osservatorio
Astrofisico di Torino, via Osservatorio 20, 10025 Pino Torinese, Italy
}

\date{Received ?? / Accepted 09/09/2016}

\abstract{
Extragalactic radio sources have been classified into two classes,  Fanaroff-Riley I and II, which differ in morphology and radio  power. Strongly emitting sources belong to the edge-brightened FR~II class,  and weakly emitting sources to the edge-darkened FR~I class. The origin of this  dichotomy is not yet fully understood. Numerical simulations are successful
  in generating FR~II morphologies, but they fail to reproduce the diffuse  structure of FR~Is. 
}{
By means of hydro-dynamical 3D simulations of supersonic jets, we investigatehow the displayed morphologies depend on the jet parameters. Bow shocks and Mach disks at the jet head, which are probably responsible for the hot spots
in the FR~II sources, disappear for a jet kinetic power ${\cal L}_{\mathrm
  kin} \lesssim 10^{43}$ erg s$^{-1}$. This threshold compares favorably with
the luminosity at which the FR~I/FR~II transition is observed. 
}{
The problem is addressed by numerical means carrying out 3D HD simulations of supersonic jets that propagate in a non-homogeneous medium with the ambient temperature that increases with distance from the jet origin, which maintains constant pressure. 
}{
The jet energy in the lower
power sources, instead of being deposited at the terminal
shock, is gradually dissipated by the turbulence. The jets spread out while
propagating, and they smoothly decelerate while mixing with the ambient medium and
produce the plumes characteristic of FR~I objects. 
}{
Three-dimensionality is an essential ingredient to explore the FR~I evolution because {  the properties of turbulence in two and three
dimensions are very different, since there is no energy cascade to small scales in two dimensions, and
two-dimensional simulations with the same parameters lead to FRII-like behavior}.
}


\maketitle

\section{Introduction}

Extragalactic radio sources have been classified into two categories \citep{FR74} based upon their radio morphology: a first class of objects, named Fanaroff-Riley I (FR~I), which is preferentially found in rich clusters and hosted by weak-lined galaxies, shows jet-dominated emission and two-sided jets at the kiloparsec scale that smoothly extend into the intracluster medium, where they form large-scale plumes or tails of diffuse radio emission. The second class, named Fanaroff-Riley II (FR~II, or classical doubles), found in poorer environments and hosted by strong emission-line galaxies, presents lobe-dominated emission
and one-sided jets at the kpc scale that abruptly terminate into hot-spots of emission.

In addition to morphology, FR~I and FR~II radio sources have
been distinguished based on power: objects below $\sim 10^{25} h_ {70}^2$ W Hz$^{-1}$ str$^{-1}$ at 178 MHz were typically found to be FR~I sources. A perhaps more illuminating criterion was found by \citet{Ledl96}, who plotted the radio luminosity at $1.4$ GHz against the optical absolute magnitude of the host galaxy: they found the bordering line of FR~I to FR~II regions to correlate as $L_R \propto L_{opt}^{1.7} $, that is, in a luminous galaxy more radio power is required to form a FR~II radio source. This correlation is important since it can be interpreted as an indication that the environment may play a crucial role in determining the source structure. Furthermore, a class of hybrid sources has been discovered that shows FR~I structure on one side of the radio source and FR~II morphology on the other \citep{gopal00}. These arguments describe the basic question of the origin of
the FR~I and FR~II dichotomy, whether the shapes are intrinsic or moulded
by the environment \citep{gopal00, wold07, kawa09, Mass03}. 

Recent studies based on the cross correlation of wide-area optical and radio surveys \citep{best12} unveiled yet another class of compact radio-galaxies representing the majority of the local radio-loud AGN population. Because they lack the prominent extended radio structures characteristic of the other FR classes, they were dubbed FR~0 \citep{baldi09, baldi10b, sadler14, baldi15}. Nonetheless, FR~0s often show radio jets, but they extend only a few kpc at most \citep{baldi15}. This suggests that in most radio-loud AGN, the jets fail to propagate to (or become too faint to be detected at) radii exceeding the size of their host.

The distorted, diffuse, and plume-like morphologies of FR~I sources led researchers to model them as turbulent flows \citep{Bicknell84, Bicknell86, Komissarov90a, Komissarov90b,  DeYoung93}, while the characteristics of FR~II, such as their linear structure and
the hot-spots at the jet termination, are associated with hypersonic flows. The difference in morphology between the two classes therefore reflects a difference in how the jet energy is dissipated during jet propagation: in the first case, the jet gradually dissipates its energy and is characterized by entrainment of the ambient material, while in the second case, the jet retains its velocity and dumps all its energy at its termination, forming the observed hot-spots.  While the dynamics of jets with high Mach number has been widely studied by means of numerical simulations in 2D and 3D
by, for example, \citet{MBF96},  \citet{zanni03}, \citet{hard13},  \citet{hard14}, 3D simulations of transonic jets have been carried out following the evolution of instabilities to turbulence \citep{bass95, hardee95, loken97, bodo98}, and simulations of turbulent jets that include the jet head propagation are {  limited to \citet{Nawaz14, Nawaz16}, who investigated the properties of the jet in Hydra A. With this paper, we therefore intend to start a systematic study of these flows}. We perform three-dimensional simulations  of the propagation of jets with low Mach number in a stratified medium,  which is meant to model the interstellar-intracluster transition. In this first paper we perform hydrodynamic simulations
and neglect the effects of the magnetic field.

 An important point to consider is that both FR~I and  FR~II radio sources show evidence of relativistic flows at the parsec scale, and therefore a deceleration to sub-relativistic velocities must occur in  FR~Is  between the inner region and the {kiloparsec} scale \citep{gg94, gg01, laing14}. Several models have been proposed for the deceleration mechanism \citep{Bicknell94, Bicknell95,  Komissarov94, Bowman96, DeYoung05} and numerical simulations of these processes have been performed \citep{Perucho07,  rossi08, Tchekhovskoy16}. In this paper, however, we assume that the deceleration has already taken place, and we consider a scale where the jet is  non-relativistic. Moreover, we stress that to study the transition to turbulence and the turbulent structure of these flows, it is essential to perform the simulations in three dimensions.
The behavior of 2D jets with the same physical parameters is  completely different from their 3D counterparts.

The plan of the paper is the following: in Sect. \ref{sec:theory} we describe the numerical setup and show the equations we solve, Sect. \ref{sec:obs} outlines the observational framework and constrains the physical parameters, and in Sect. \ref{sec:results} we present and discuss the obtained results. The conclusions are drawn in Sect. \ref{sec:conclusions}.

\section{Numerical setup} 
\label{sec:theory}
%
%
%

\subsection{Equations and method of solution} 
\label{sec:mhd}

For our purposes, we solve the Euler equations of gas dynamics, which can be written in quasi-linear form as
\begin{equation}\label{eq:continuity}
\frac{\partial\rho}{\partial t} + \nabla \cdot (\rho \vec v) = 0\,,
\end{equation} 

\begin{equation}\label{eq:euler} 
\frac{\partial\vec v}{\partial t} + (\vec v \cdot \nabla)\vec v = - \frac{1}{\rho}\nabla P, 
\end{equation} 

\begin{equation} \label{eq:energy}
\frac{\partial P}{\partial t} + \vec v \cdot \nabla P + \Gamma P \nabla \cdot \vec v = 0 \,,
\end{equation} 

\begin{equation}\label{eq:tracer}
\frac{\partial f}{\partial t} + \vec v \cdot \nabla f = 0\,. 
\end{equation} 
The quantities $\rho$, $P$, and $\vec v$ are the density, pressure, and velocity, respectively, and $\Gamma = 5/3$ is the ratio of the specific heats.
The jet and the external material are distinguished using a passive tracer, $f$, set equal to unity for the injected 
jet material and equal to zero for the ambient medium. We{  note that the choice of $\Gamma = 5/3$ is consistent with a
supersonic nonrelativistic jets with Mach numbers > 4 (see discussion below).}

Equations (\ref{eq:continuity})-(\ref{eq:tracer}) were solved using the piecewise parabolic method (PPM) and the HLLC Riemann solver of the PLUTO code (Mignone et al 2007, 2012).

\subsection{Initial and boundary conditions}

The 3D simulations were carried out on a Cartesian domain with coordinates in the range $x\in [-L/2,L/2]$, $y\in [0,L_y]$ and $z\in [-L/2,L/2]$ (lengths are expressed in units of the jet radius $r_j$ and  $y$ is the direction of jet propagation).
At $t=0$, the domain is filled with a perfect gas at uniform pressure but spherically stratified according to a King-like profile in density:
\begin{equation}
\rho(r)=\frac{1}{\eta}\frac{1}{1+\left(r/r_{\mathrm c}\right)^\alpha} \label{eq:king} ,
\end{equation}
where the density is expressed in units of the jet density $\rho_j$, $r=\sqrt{x^2+y^2+z^2}$ is the spherical radius, $r_{\mathrm c}$ is the core radius, and $\eta$ is the ratio $\rho_j/\rho_c$ between the jet density and the core density. We set  $r_{\mathrm c}=40 r_j$ throughout and considered different values of the parameters $\alpha$ and $\eta$. 
The ambient temperature increases  with radius as
\begin{equation}
T \propto 1+\left(r/r_{\mathrm c}\right)^\alpha
\end{equation}
to maintain the pressure uniform.

We imposed zero-gradient boundary conditions on all computational boundaries with the exception of the injection region located at $y=0$.
Here we prescribed inside the unit circle a constant cylindrical inflow directed along the $y$ direction, and in all but one case (see Table \ref{labvalues}), in pressure equilibrium with the ambient.
The inflow jet values are given by 
\begin{equation} \label{eq:inflow} 
  \begin{array}{lcl}
  \rho_j     &=&  1         \,, \\  \noalign{\medskip}
  v_{y,j}    &=&  M         \,, \\  \noalign{\medskip} 
  p_j        &=&  1/\Gamma  \,, \\  \noalign{\medskip}
  f_j        &=&  1    \,,
  \end{array}
\end{equation} 
where velocity is measured in units of the jet sound speed on the axis, $c_{\mathrm sj}$, therefore $M$ represents the jet Mach number, and pressure is measured in units of $\rho_j c_{sj}^2$.
Outside the jet nozzle, reflective boundary conditions hold.
To avoid sharp transitions, we smoothly joined the injection and reflective boundary values in the following way:
%
%
\begin{equation} \label{eq:profiles}
  Q(x,z,t)  = Q_r(x,z,t) + \frac{Q_j - Q_r(x,z,t)}{\cosh\left[(R/R_s)^n\right]}
,\end{equation}
where $Q = \{\rho,\, v_x,\, \rho v_y,\, v_z,\, p,\, f\}$ are primitive flow quantities with the exception of the jet velocity, which was replaced by the $y-$momentum, while $R =  \sqrt{x^2+z^2}$  is the cylindrical radius.
We note that $Q_r(x,z,t)$ are the corresponding time-dependent reflected values, while $Q_j$ are the constant injection values given by Eqs. (\ref{eq:inflow}).
In Eq. (\ref{eq:profiles}) we set $R_s=1$ and $n=6$ for all variables except for the density, for which we used $R_s=1.4$ and $n=8$.
This choice ensured monotonicity in the first $(\rho v_y)$ and second $(\rho v^2_y)$ fluid outflow momenta \citep{MBF96}.
We did not explicitly perturb the jet at its inlet, differently from \cite{Mig2010}. The growing non-axially symmetric modes originate from numerical noise.

The physical domain was covered by $N_x \times N_y \times N_z$ computational zones that are not necessarily uniformly spaced. For domains with a large physical size, we employed a uniform grid resolution in the central zones around the beam (typically for $ |x|, \ |z| \leq 6$) and a geometrically stretched grid elsewhere. {  In the central zones we have a resolution of 10 grid points per jet radius for the cases of $M=4, 10,$ and 13 for the case of $M=40$. As shown in \cite{anjiri14}, a resolution of 12 points per jet radius produces results that agree well, within a few percent, with computations performed by doubling this resolution. The comparison between different resolutions was carried out considering both the morphological and the energy budget aspects.} The complete set of parameters for our simulations is given in Table \ref{labvalues}.

\begin{table*}[ht] 
\begin{center} 
\caption{Parameter set used in the numerical simulations.} 
\begin{tabular}{lccccccccll} 
\hline
1 & 2 & 3 & 4 & 5 & 6 & 7 & 8 & 9 & 10 & 11\\
\hline 
  & 
$\eta$ &
$M$& 
$\alpha$&   
$P_{\mathrm j} / P_{\mathrm c}$&  
$v_j$ (cm/s)&
${\cal L}_{\mathrm kin}$ (erg/s)& 
$L_x \times L_y \times L_z$& 
$N_x \times N_y \times N_z$&  
Notes & 
FR\\ 
\hline 
A & $10^{-2}$ &  4  & 0 &  1 & $5.1 \times 10^8$ & $1.1 \times 10^{42}$ & $64 \times 120 \times 64$ & $512\times 1280 \times 512$ &  Not stratified    &  I  \\
B & $10^{-2}$ &  4  & 2 &  1 & $5.1 \times 10^8$ & $1.1 \times 10^{42}$ & $64 \times 120 \times 64$ & $512\times 1280 \times 512$ &  Reference case    &  I  \\ 
C & $10^{-2}$ &  4  & 1 &  1 & $5.1 \times 10^8$ & $1.1 \times 10^{42}$ & $64 \times 120 \times 64$ & $512\times 1280 \times 512$ &  Different $\alpha$&  I  \\ 
D & $10^{-2}$ &  4  & 2 & 10 & $1.6 \times 10^9$ & $3.5 \times 10^{42}$ & $64 \times 120 \times 64$ & $512\times 1280 \times 512$ &  Overpressured     &  I  \\ 
E & $10^{-2}$ & 40  & 2 &  1 & $5.1 \times 10^9$ & $1.1 \times 10^{45}$ & $60 \times 240 \times 60$ & $576\times 2560 \times 576$ &  Highest power     &  II \\ 
F & $10^{-2}$ & 10  & 2 &  1 & $1.3 \times 10^9$ & $1.7 \times 10^{43}$ & $64 \times 120 \times 64$ & $512\times 1280 \times 512$ &  Interm. power     &  II \\ 
G & $10^{-3}$ &  4  & 2 &  1 & $1.6 \times 10^9$ & $3.5 \times 10^{42}$ & $64 \times 120 \times 64$ & $512\times 1280 \times 512$ &  Lighter jet       &  I  \\ 
\hline 
\end{tabular} 
\label{labvalues} 
\end{center}

Column description: 1) case identifier, 2) density ratio $\eta$, 3) Mach
number $M$, 4) exponent $\alpha$ characterizing the initial density
distribution, 5) ratio between jet and external pressure, 6) jet velocity, 7)
jet kinetic power, 8) domain extension in the three directions, 9) number of
grid points in the three directions, 10) short description of the case, 11)
resulting FR morphology. Jet velocity and kinetic power are computed with the
reference values for the external medium parameters (see Sect.
\ref{sec:obs}).

\end{table*}

        \section{Observational background and parameter constraints} 
\label{sec:obs}

To obtain physical values from the results of our simulations, we set  constraints derived from observational data to the values of the fundamental units of length $r_j$,  density $\rho_j$, and velocity $c_{sj}$. 

As discussed in the introduction, FR~I  jets propagate at relativistic velocities close to the central AGNs at the parsec scales, while they are non-relativistic or weakly relativistic at the
kiloparsec scales \citep{gg94, gg01}. This means that a jet-braking process must have taken place between the inner and outer scales \citep{rossi08, laing14}. Our goal here is not to study this deceleration process, however, but to consider the 
 jet propagation at scales starting from a few hundred pc, where the jet is already non-relativistic,  to the typical linear size of FR~I radio sources, which range from about 10 up to about hundred  of kiloparsec (\cite{parma87} for the B2 catalog sources). 
 
 At these spatial scales, the jet propagates in the galactic interstellar medium for the first few kiloparsec before exiting into the intracluster medium. The jet radius in most FR~I sources is at the limits of the angular resolution of radio interferometers such as the VLA, $\sim 0.1-0.3^{\prime
  \prime}$ \citep{parma87}. We assumed a fiducial value for the jet radius of $r_{\mathrm j}=100$ pc. {  A consistent limit on the jet radius is also determined by requirements on numerical resolution:  since we require the grid physical extension to be about ten kiloparsec, the jet radius cannot be much smaller than 100 pc with the given grid size if we aim to cover it with a number of points not much smaller than 10. In the following, the values of the spatial coordinates are given in kiloparsec within this assumption for the initial jet radius.}

The typical galactic core radius $r_{\mathrm c}$ attains values of about a few kiloparsec, and we assumed $r_{\mathrm c}=4$ kpc, which means 40 jet radii. X-ray observations have shown \citep{pos13} that within the galactic core, the average temperature of the ambient medium is $T_{\mathrm c} \sim 0.1 - 0.3$ keV, with a particle density of $n_{\mathrm c} \sim 1 $ cm$^{-3}$ \citep{balm08}. As the jet enters the intragroup/intracluster medium, the ambient gas is hotter by about one order of magnitude \citep{sun09, connor14} and the density drops by approximately the same amount. This is consistent with assuming the parameter $\alpha$  in Eq. \ref{eq:king}
to be between 1 and 2. 

The jet parameters can be expressed as functions of the external density, $\rho_c$, the external temperature, $T_c$,  the Mach number M, the density ratio $\eta,$ and the ratio between jet and ambient pressure as follows:

\begin{equation}
\rho_j = 0.01 \left( \frac{\eta}{0.01} \right) \, \left( \frac{\rho_c}{1 \rm{cm}^{-3}} \right) \, \rm{cm}^{-3} \,, 
\end{equation}

\begin{equation}
v_j = 5.1 \times 10^8 M \, \left( \frac{T_c}{0.2 keV} \right)^{1/2}  \, \left( \frac{\eta}{0.01} \right)^{-1/2} \,  \left(  \frac{P_j}{P_c}\right)^{1/2} \rm{cm}\, {s}^{-1}  \,.\end{equation}
We can also derive the jet kinetic power as

\begin{equation} 
\begin{split}
{\cal L}_{\mathrm kin}= 1.1 \times 10^{42} \left( \frac{r_j}{100 pc}\right)^2 \, \left( \frac{\rho_c}{1 \rm{cm}^{-3}} \right) \,
 \left( \frac{T_{c}}{0.2 keV}  \right)^{3/2} \\
 \left(\frac{M}{4}\right)^3 \, \left( \frac{\eta}{0.01} \right)^{-1/2}\,  \left(  \frac{P_j}{P_c}\right)^{3/2}\rm{erg} \, \rm{s}^{-1}.
\end{split}
\end{equation} 

The computational time unit $\tau$ is the sound travel time over the initial jet radius, which in physical units corresponds to
\begin{equation}
\begin{split}
\tau = \frac{r_j} {c_{sj}}= 7.7 \times 10^4 \left( \frac{r_j}{100 pc} \right)   \, \left( \frac{T_c}{0.2 keV} \right)^{-1/2}  \\
 \left( \frac{\eta}{0.01} \right)^{1/2} \,\left(  \frac{P_j}{P_c}\right)^{-1/2} \, yrs \,.
\end{split}
\end{equation}
Our typical simulation covers about 100 to 1000 time units, corresponding to about $10^7 - 10^8$ yrs of the source lifetime, while it has traveled more than 10 kpc in the ambient medium.

Various methods for converting the kinetic jet power ${\cal L}_{\mathrm {kin}}$ into
observed radio luminosity $L_{\mathrm r}$ have been proposed; we are here
mainly interested in an estimate of ${\cal L}_{\mathrm {kin}}$ corresponding to
the power at which the FR~I/FR~II transition occurs, ${\cal L}_{\mathrm
  trans}$. \cite{willott99} obtained a relation according to which
${\cal L}_{\mathrm {kin}}$ scales as $L_{\mathrm r}^{6/7}$, leading to ${\cal L}_{\mathrm
  trans} \sim 1.4 \times 10^{43}$ erg s$^{-1}$. \cite{birzan04} estimated
${\cal L}_{\mathrm {kin}}$ from the observations of the X-ray cavities inflated by
radio AGN. The value of ${\cal L}_{\rm trans}$ based on their calibration support the
above estimate, returning values in the range ${\cal L}_{\mathrm trans} \sim 3 \times
10^{42} - 5 \times 10^{43} $ erg s$^{-1}$ at the FR~I/FR~II power transition.

We performed various simulations that we describe in the following section,
starting at ${\cal L}_{\mathrm {kin}} = 1.1 \times 10^{42}$ erg s$^{-1}$, within the
expected range of FR~I power, and then increased the jet power into the FR~II
regime.

\section{Results and discussion} 
\label{sec:results}

We have carried out a series of simulation runs to explore the effects of the different parameters on the jet propagation, evolution, and resulting morphologies. 

\begin{figure*}[!ht] 
\begin{center} 
\includegraphics[width=1.4\columnwidth]{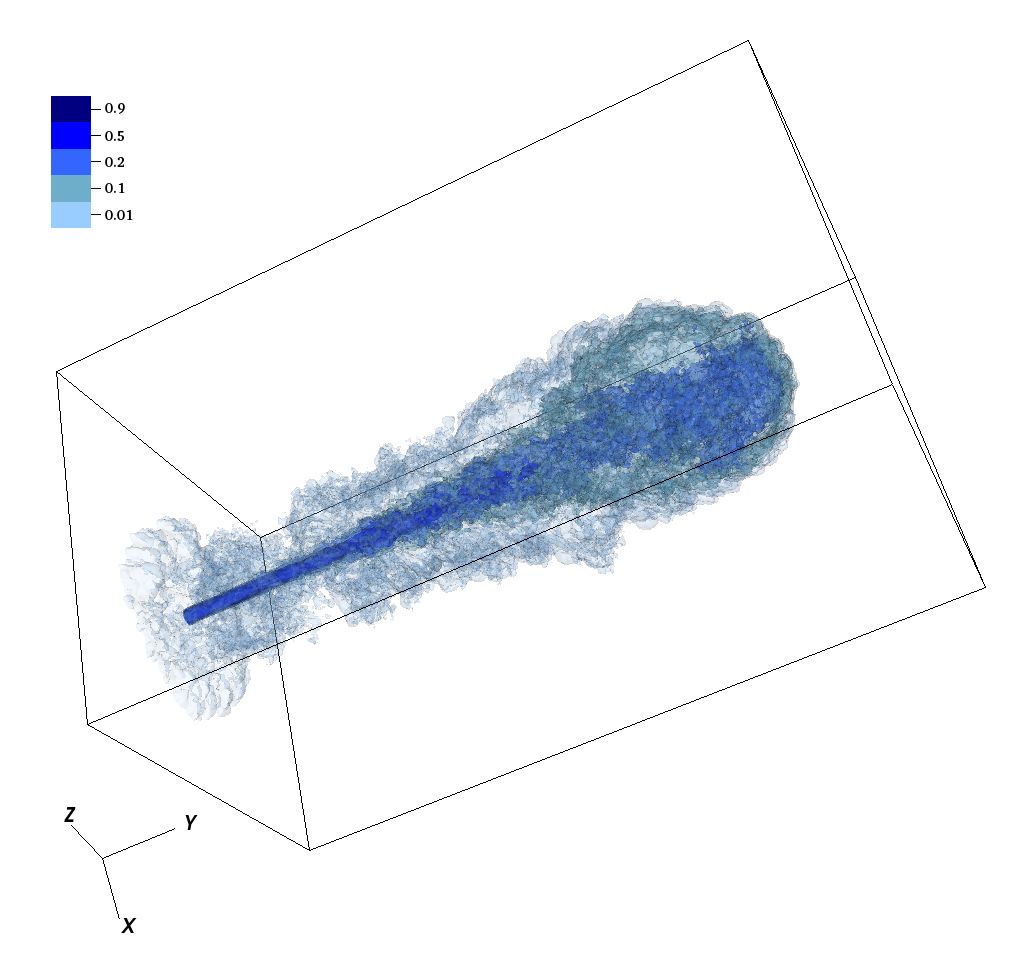} 
\end{center} 
\caption{3D isocontours of the tracer distribution for case A ($M=4$, $\eta = 10^{-2}$, ${\cal L}_{\mathrm
    {kin}} = 1.1 \times 10^{42}$ erg s$^{-1}$) at $t=740$ time units, $5.7\times
  10^7$ yrs. The size of the computational box is $6.4 \times 12 \times 6.4$ kpc.
}
\label{fig3D} 
\end{figure*}

Although we simulated jets propagating in a stratified medium,
which reproduces
the real ambient medium of radio galaxies, we started our analysis with an
initial simulation (case A in Table 1) in which the ambient density was
constant. This was motivated by the need of comparing our results with those
obtained in previous works. The jet has a Mach number $M=4$ and a density
contrast of $\eta=10^{-2}$ corresponding to ${\cal L}_{\mathrm {kin}} = 1.1
\times 10^{42}$ erg s$^{-1}$, well below the FR~I/FR~II power transition.  The
outcome of the simulation is presented in the 3D image shown in
Fig. \ref{fig3D}. After an initial collimated phase, the jet disrupts and
expands into the ambient medium. We present an image of the jet tracer
and not of emission, but it can be envisaged that this diffuse gas
distribution will give rise to an FR~I radio morphology.

\begin{figure*}[!ht] 
\begin{center} 
\includegraphics[width=1.2\columnwidth]{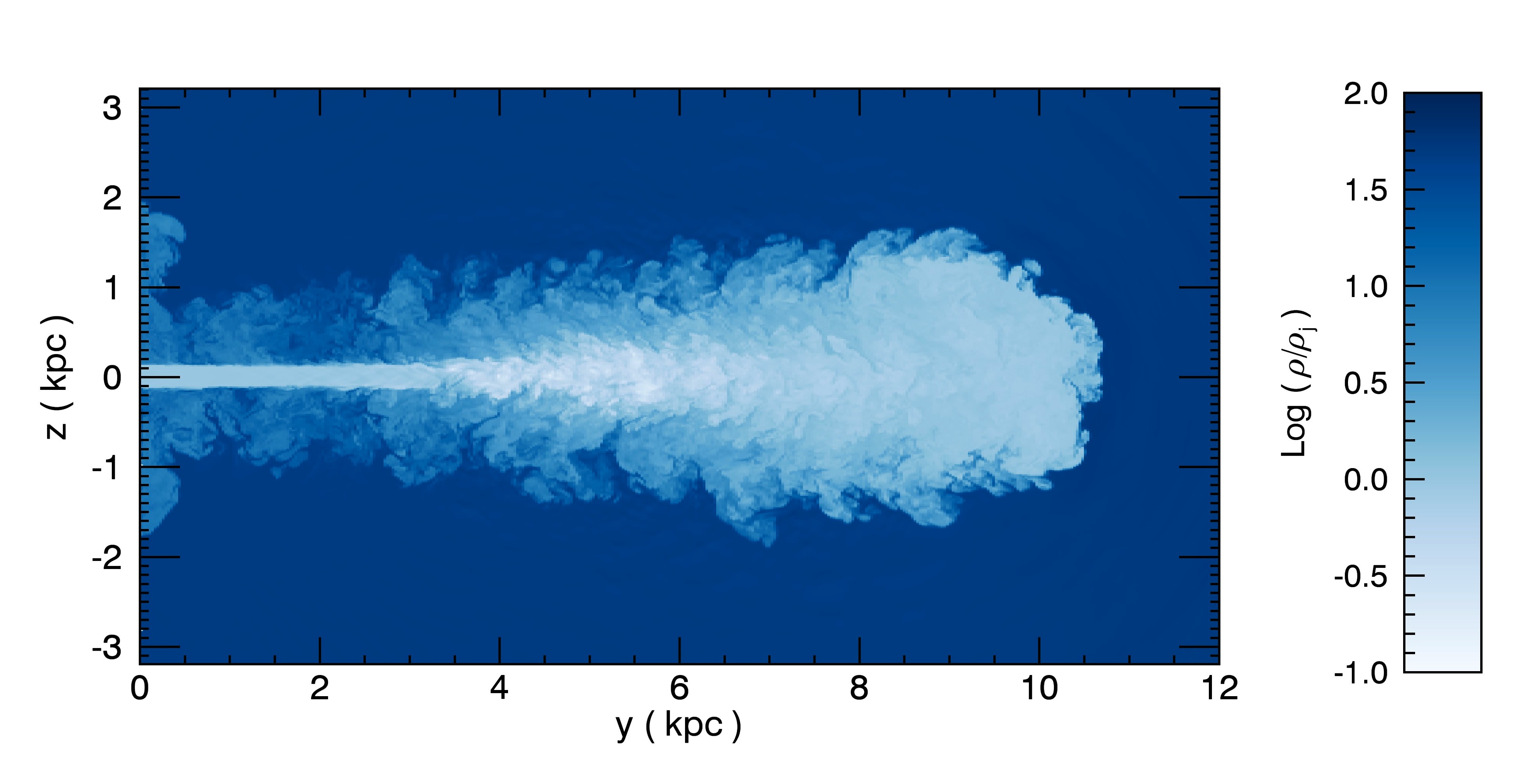}
\includegraphics[width=0.75\columnwidth]{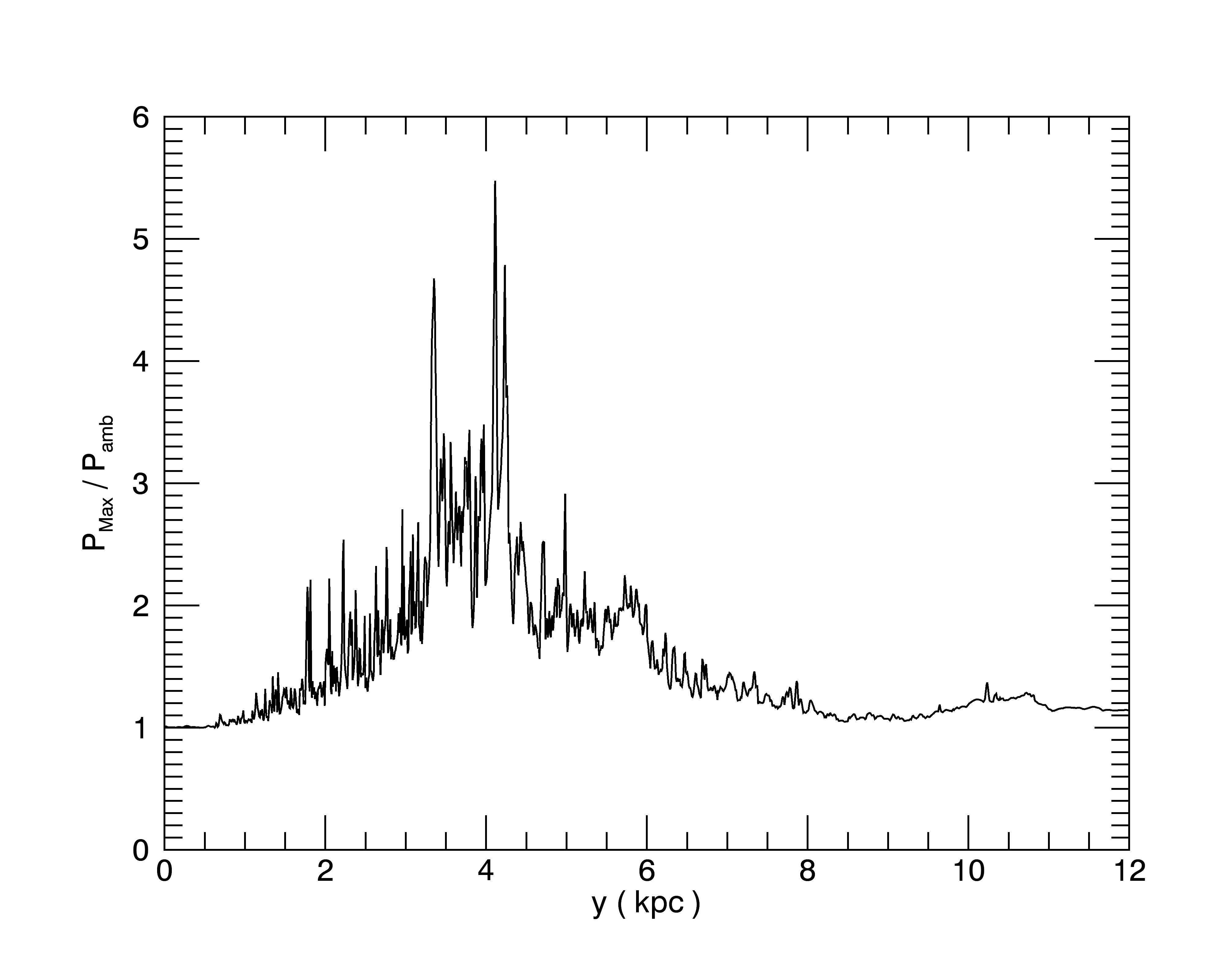}
\end{center} 
\caption{Results for case A ($M=4$, $\eta = 10^{-2}$, ${\cal L}_{\mathrm
    {kin}} = 1.1 \times 10^{42}$ erg s$^{-1}$) at $t=740$ time units, $5.7\times
  10^7$ yrs. Left: cut in the $(y,z)$ plane of  the
  logarithmic density distribution in units of jet density. The spatial units are in jet radii (100 pc) and the image extends
  over 12 kpc in the jet direction and over 6 kpc in the transverse direction. 
  Right: maximum pressure 
on a $xz$ plane at a given position $y$ along the jet as a function of $y$. The pressure is plotted, from now on, in
units of the ambient value (instead of the {\it computational} units $\rho_j  c_{sj}^2$).
}
\label{fig:A} 
\end{figure*}

This is confirmed by Fig. \ref{fig:A}, in which we show the maximum pressure 
on a transverse ($xz$) plane at a given position $y$ along the jet as a function of $y$.
 The pressure rises along the jet, reaching
its maximum at $r \sim 40,$ and then it steadily decreases. At the jet head there is
only a small increase, $\sim$20\%, indicating a weak shock.

\begin{figure*}[!ht] 
\begin{center} 
\includegraphics[width=1.2\columnwidth]{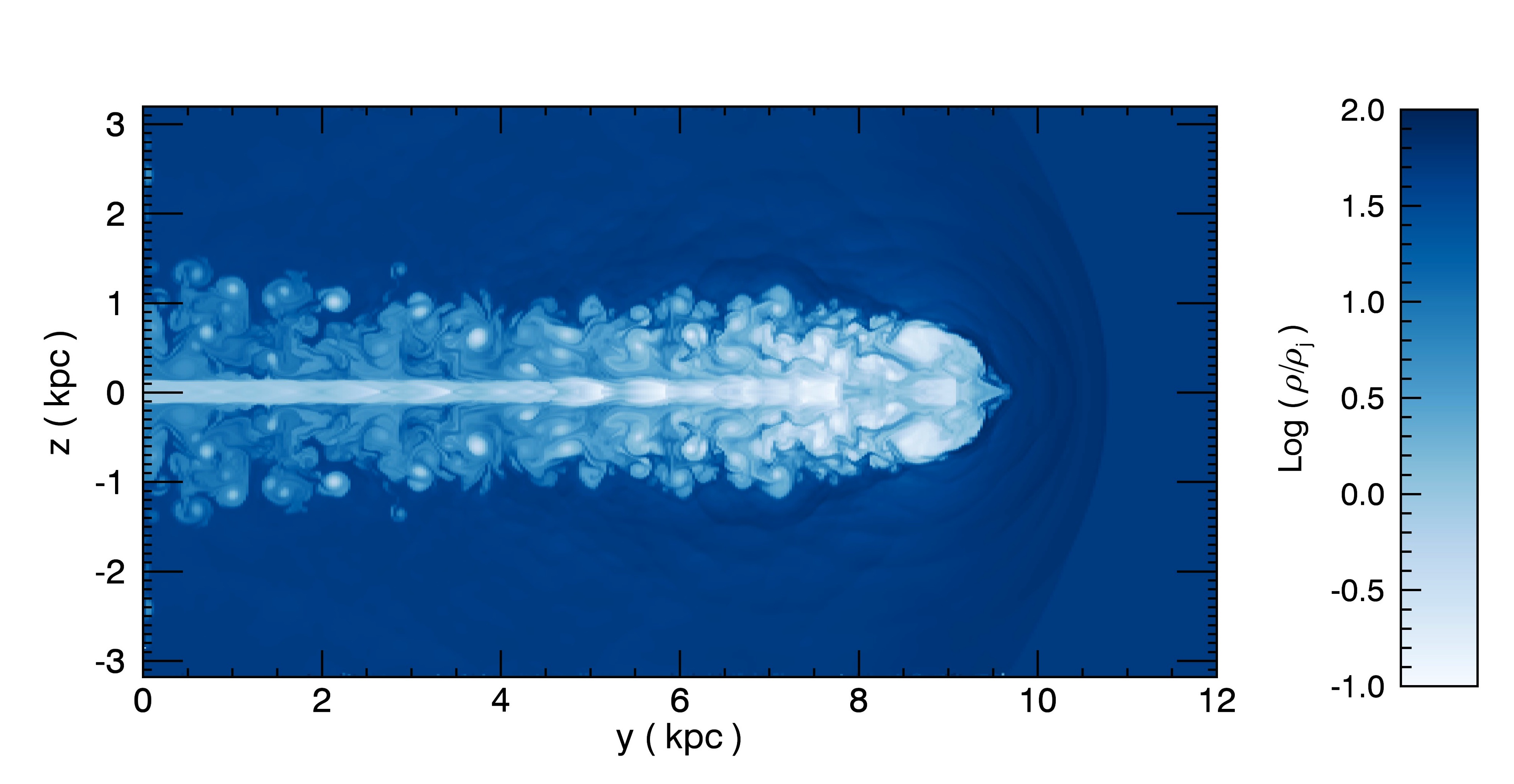} 
\includegraphics[width=0.75\columnwidth]{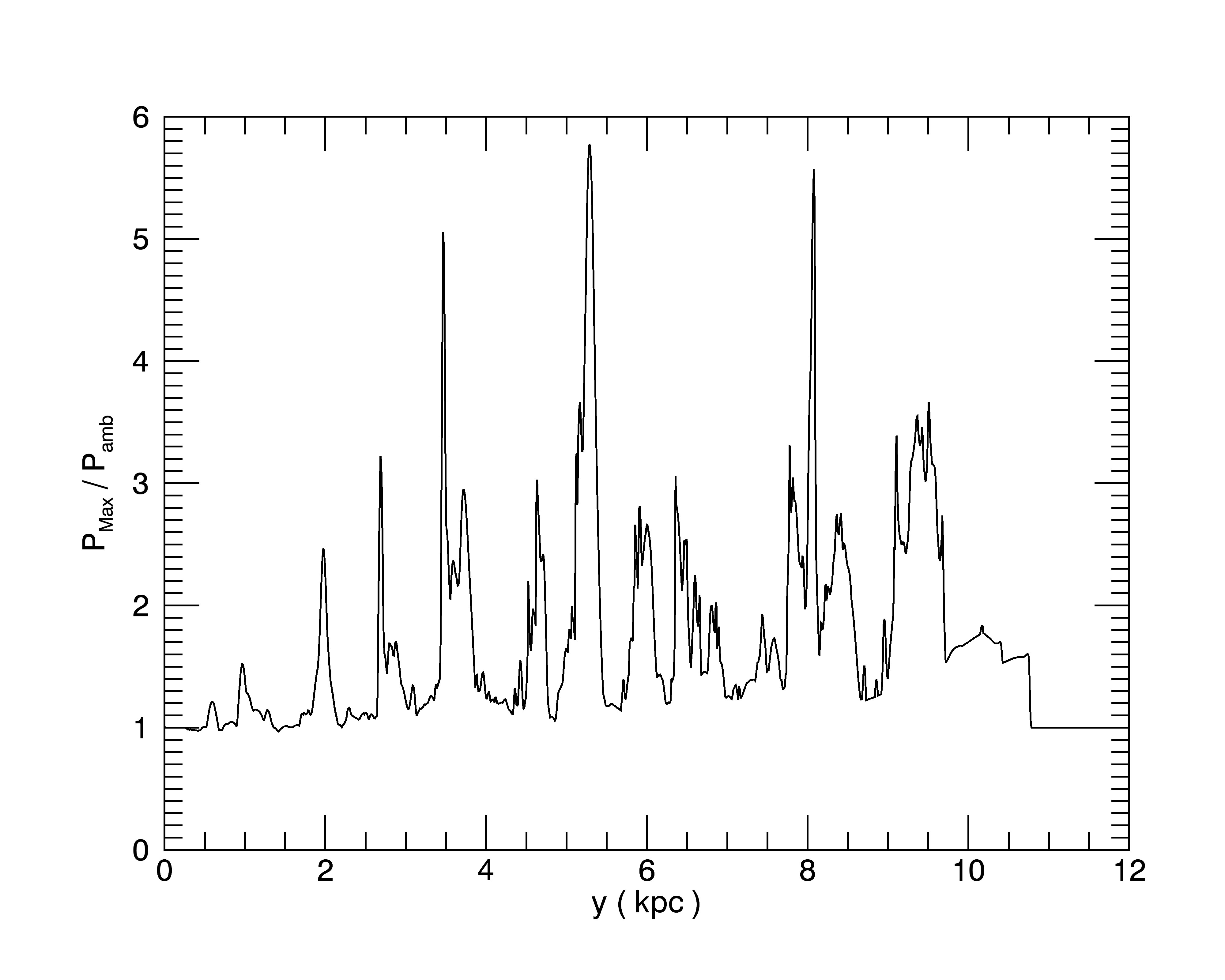} 
\end{center} 
\caption{Results for a 2D simulation with the same jet parameters as case A and at $t=500$ time units, $4 \times 10^7$ yrs.
}
\label{Fig:2D} 
\end{figure*}

We simulated a jet described by the same parameters, but in 2D {  cylindrical geometry (axisymmetric)}, see
Fig. \ref{Fig:2D}, and the results are radically different. The jet remains
collimated over its whole length.  The pressure displays several local maxima
associated with the jet internal shocks and, more importantly, with its
head.

{  This can be explained by recalling that the properties of turbulence in two and three dimensions 
are very different because in two dimensions we do not have the energy cascade to small scales. 
Consequently, the entrainment properties and the jet evolution become strongly diversified.}  In particular, the FR~I
morphology resulting from the propagation of a low-power jet arises and can be
explored only with high-resolution 3D simulations.  To extract
physical information from our simulations in a more realistic framework, all
the cases that follow were carried out in the stratified medium described in
Sect. 2.

\begin{figure*}[!ht] 
\begin{center} 
\includegraphics[width=1.6\columnwidth]{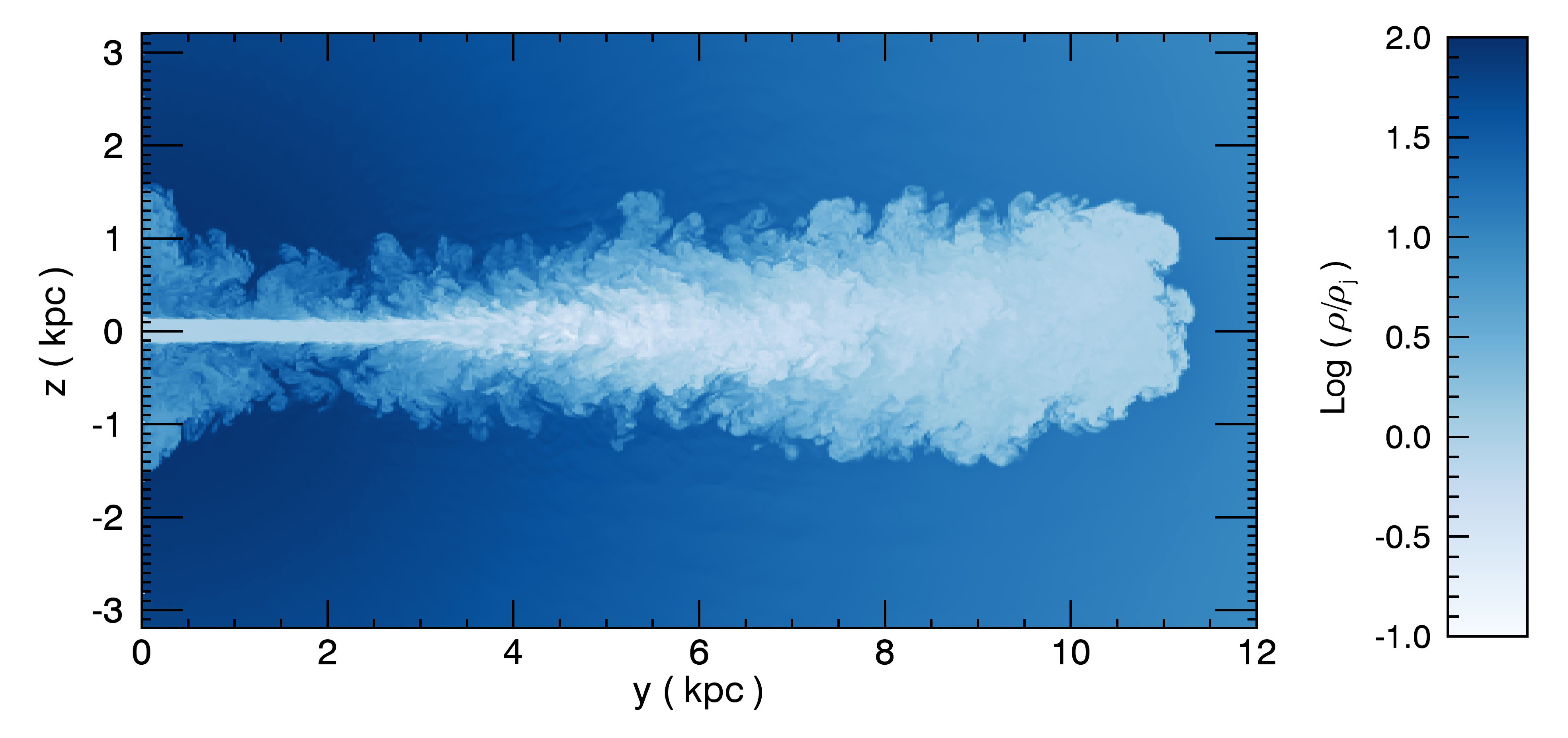}
\includegraphics[width=1.6\columnwidth]{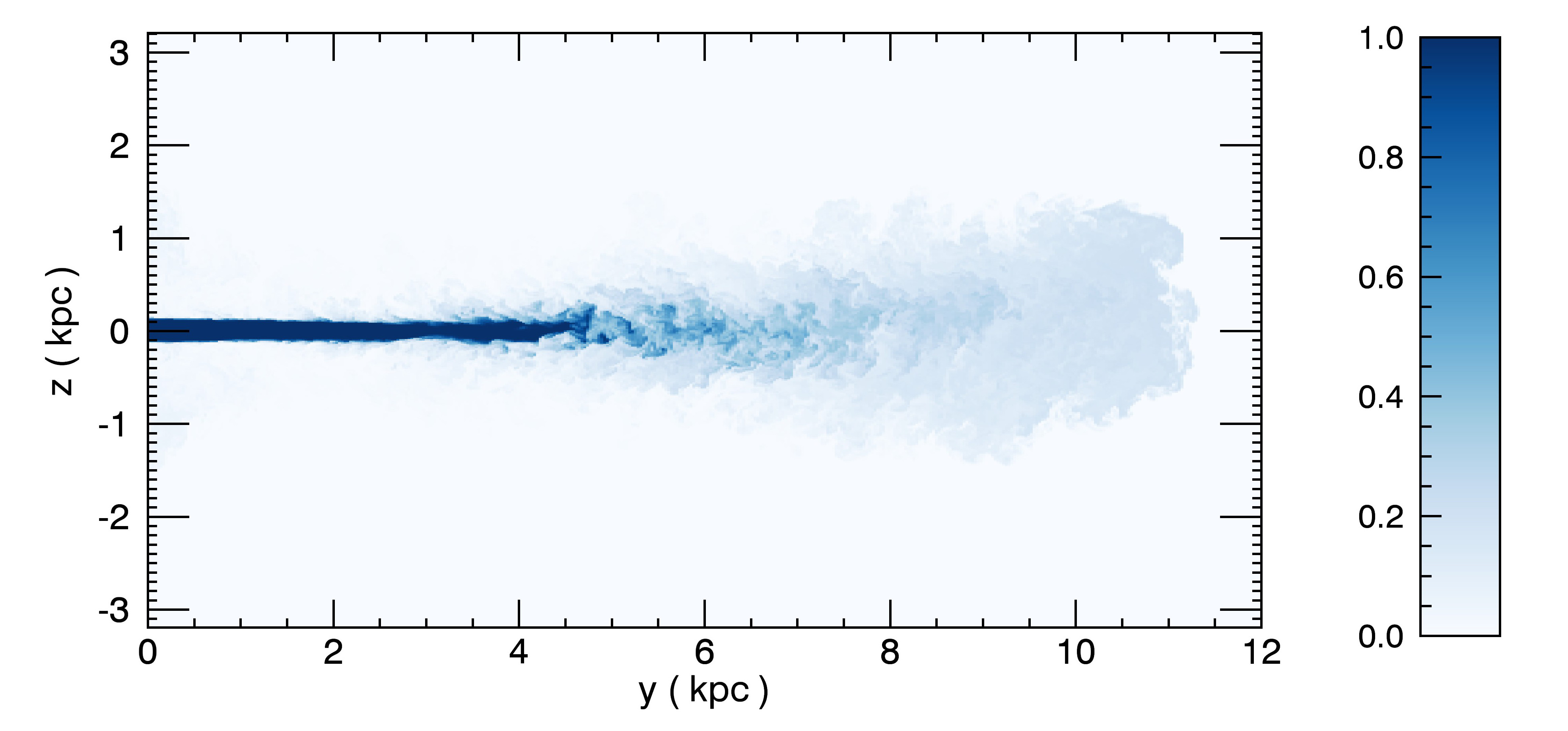}
\vspace{-0.3in}
\includegraphics[width=1.6\columnwidth]{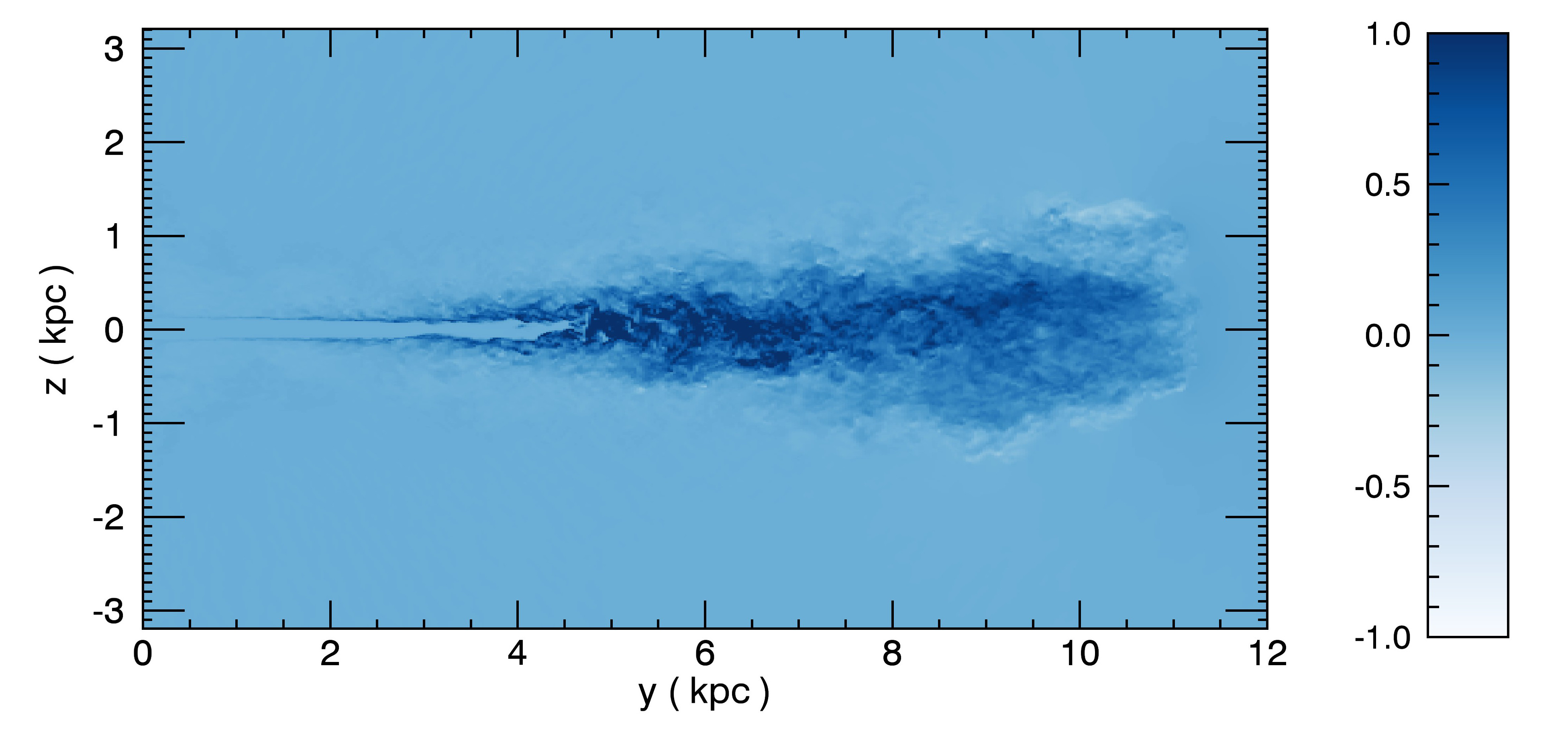} 
\end{center} 
\vspace{.1in}
\caption{Results for our reference case B ($M=4$, $\eta = 10^{-2}$, ${\cal
    L}_{\mathrm kin} \sim 1.1 \times 10^{42} \ {\mathrm ergs \ s}^{-1}$), at
  $t=640$ time units, $4.9\times 10^7$ yrs). The spatial units are in jet radii
  (100 pc) and the image extends over 12  kpc in the jet direction and over 6 kpc in the transverse 
  direction. Cuts in the $(y,z)$
  plane of (top) the logarithmic density distribution in units of jet density,
  (center) tracer distribution and (bottom) distribution of the longitudinal
  velocity of the external medium in units of  $c_{sj}$.
}
\label{fig:B} 
\end{figure*}

In the top panel of Fig. \ref{fig:B} we show a cut in the $(y,z)$ plane of the
logarithmic density distribution for case B of Table \ref{labvalues}
($M=4$, $\eta=10^{-2}$, the same parameters as for case A, but including the
external density stratification) at $5.2\times 10^7$ yrs of the
jet lifetime and at a traveled distance of 12 kpc. The general features are very similar to
those shown by case A, without ambient gas stratification. No sign of the head Mach disk  is observed, and the jet smoothly mixes with the
ambient medium.  This is confirmed by the cuts of the tracer distribution and
of the longitudinal velocity of the matter belonging to the external medium, {  that is, the quantity $(1-f) \times v_y$} in the
central and bottom panels of Fig. \ref{fig:B}. {  We also note that the bow shock has disappeared from the 
ambient medium  since the head propagation has become subsonic at these distances.} The jet remains well collimated
out to 30-40 jet radii, then it rapidly widens at large distances. This is the same value of the core radius of   the external gas distribution; however, this is observed also in case A,   where no stratification is present, and it should be considered as a   coincidence. This distance is most likely related to the growth length of instabilities that cause
  the transition to turbulence. At the same
location it decelerates and entrains external material that is also globally
accelerated. This behavior is preserved qualitatively out to the jet head,
with just a further widening and deceleration.

\begin{figure*}[!ht] 
\begin{center}
\includegraphics[width=1.6\columnwidth]{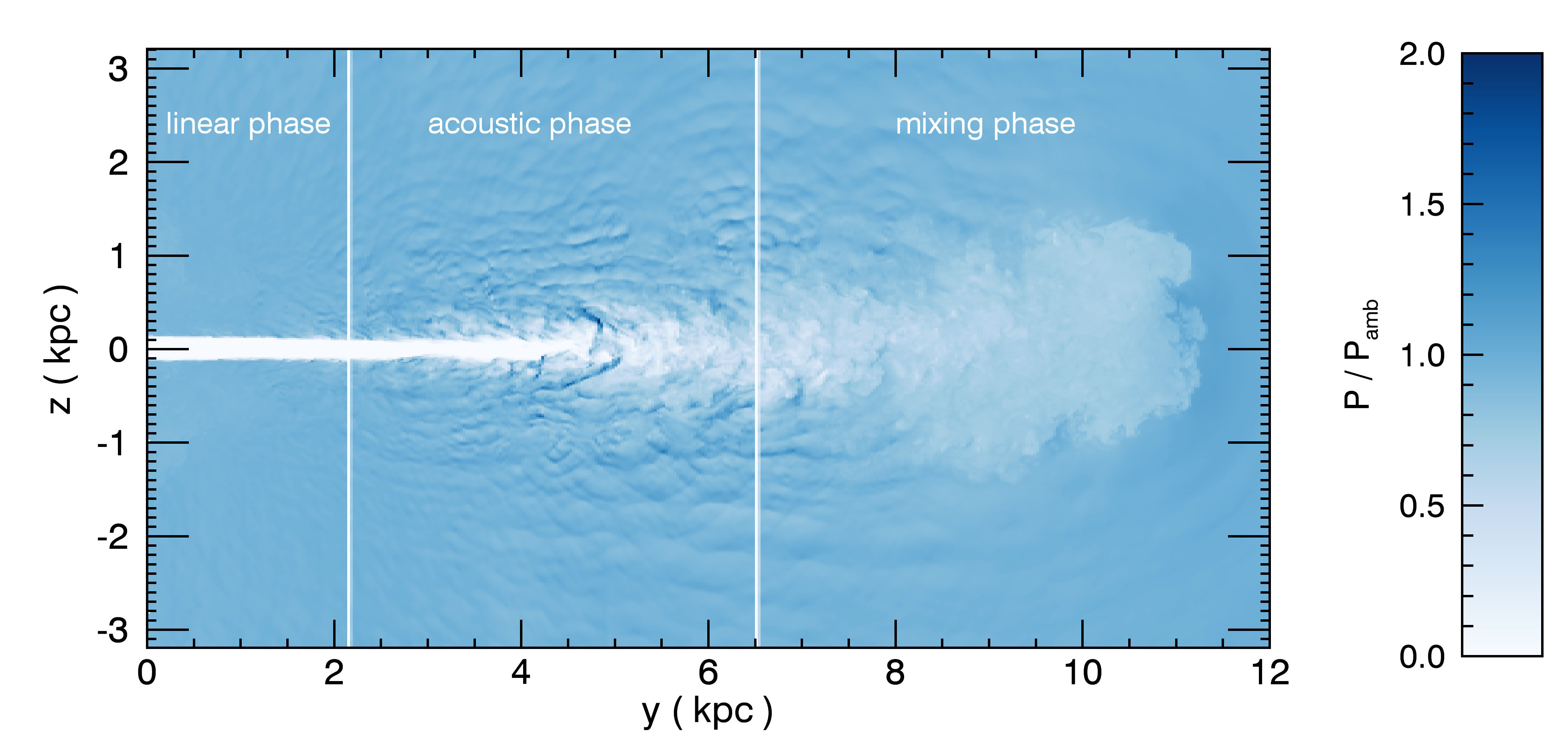} 
\includegraphics[width=1.6\columnwidth]{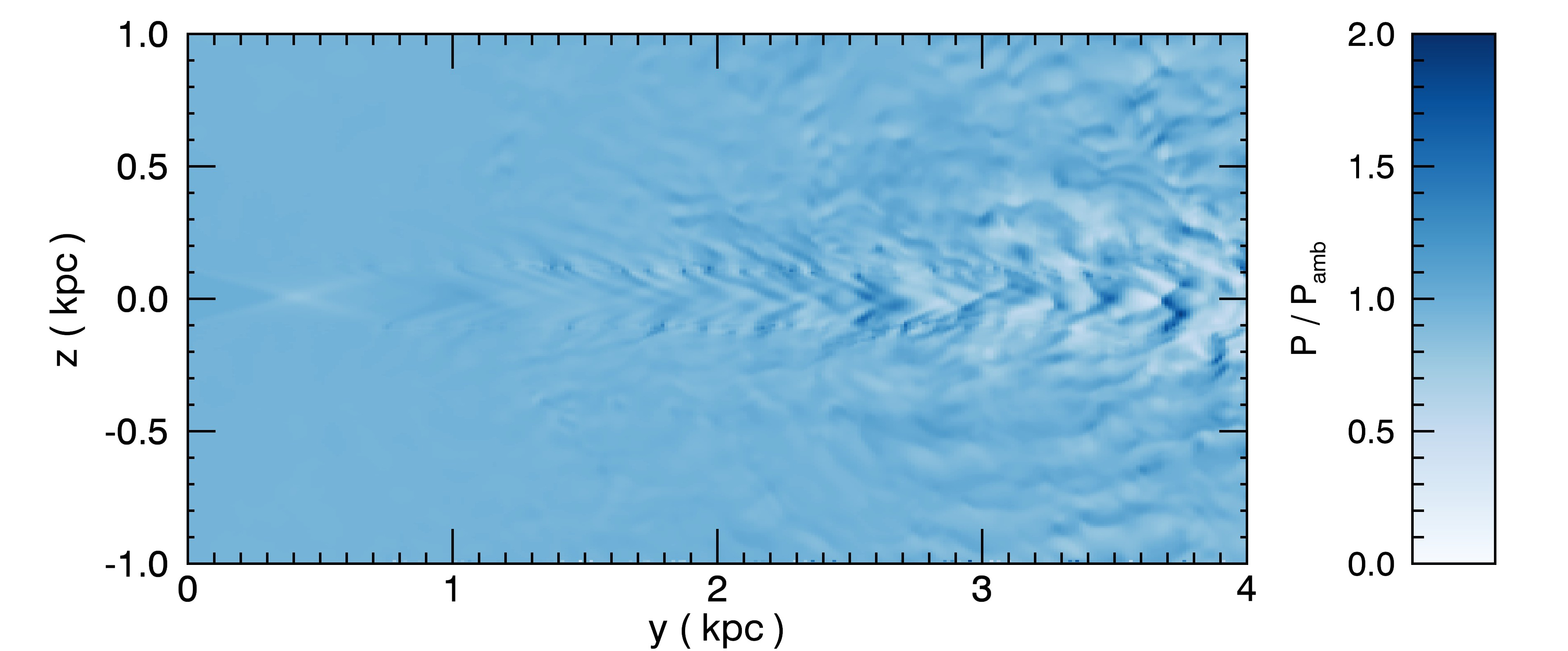} 
\end{center} 
\caption{Same as in Fig.
\ref{fig:B}, but for the pressure distribution of the external medium. The three phases of the evolution are labeled (top panel). {  The bottom panel shows the total pressure distribution in the initial part of the jet with internal shocks.}} 
\label{fig:phases_A}
\end{figure*}

As first pointed out by \cite{bodo94}, who studied the nonlinear temporal evolution of jet{  velocity shear instabilities (Kelvin-Helmholtz instabilities) }in 2D {  cylindrical symmetry,} and then by \cite{bodo98} in the 3D extension of this investigation, the evolution of unstable modes
follows a trend that is essentially unchanged in the general features and that
develops in three phases: 1) a first {\it linear phase} where the unstable
modes grow according to the linear theory until internal shocks start to form;
2) this is followed by an {\it acoustic phase} where the growth of the
internal shocks is accompanied by a global deformation of the jet, which
drives shocks into the external medium; these shocks carry both momentum and
energy away from the jet and transfer them to the external medium; 3) eventually a final {\it mixing phase} where, as a consequence of the shock
evolution, turbulent mixing between jet and the ambient medium begins. In Fig. \ref{fig:phases_A} {  (top panel)} the cut of the pressure distribution of the {\it external}
medium, i.e. $(1-f) \times P$, is suitable to show how these three phases develop in space. {  In Fig. \ref{fig:phases_A} (bottom panel) we show a detail of the total pressure in the central and initial section of the domain that is relevant for the linear and, partly, the acoustic phase. Oblique shocks internal to the jet are observed.}

\begin{figure*}[!ht] 
\begin{center} 
\includegraphics[width=1.6\columnwidth]{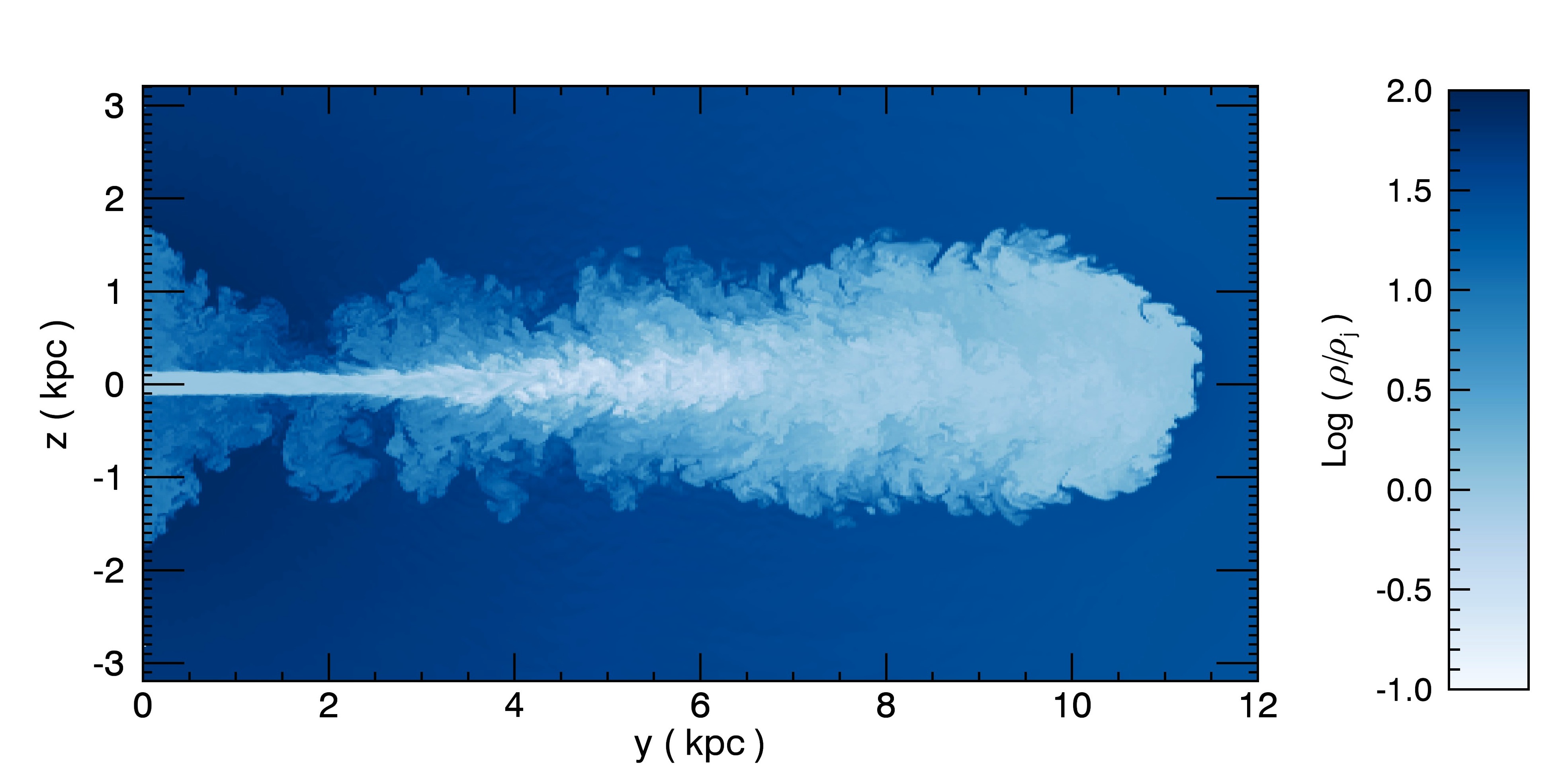} 
\end{center} 
\caption{Cut in the $(y,z)$ plane of the
logarithmic density distribution for case C, differing from the reference case B for the 
value of $\alpha$ that describes the external medium, at $t=740$ time units, $5.7 \times 10^7$ yrs.} 
\label{fig:C} 
\end{figure*}

The main properties obtained for case B are also reproduced in case C, which
differs only for the parameter $\alpha$ , which describes the density profile of the
ambient medium (see Fig. \ref{fig:C}). We only note a slightly slower
(by $\sim$ 10\%) advance of the jet head, which is due to the slower density decrease of
the external gas.

\begin{figure*}[!ht] 
\begin{center} 
\includegraphics[width=1.6\columnwidth]{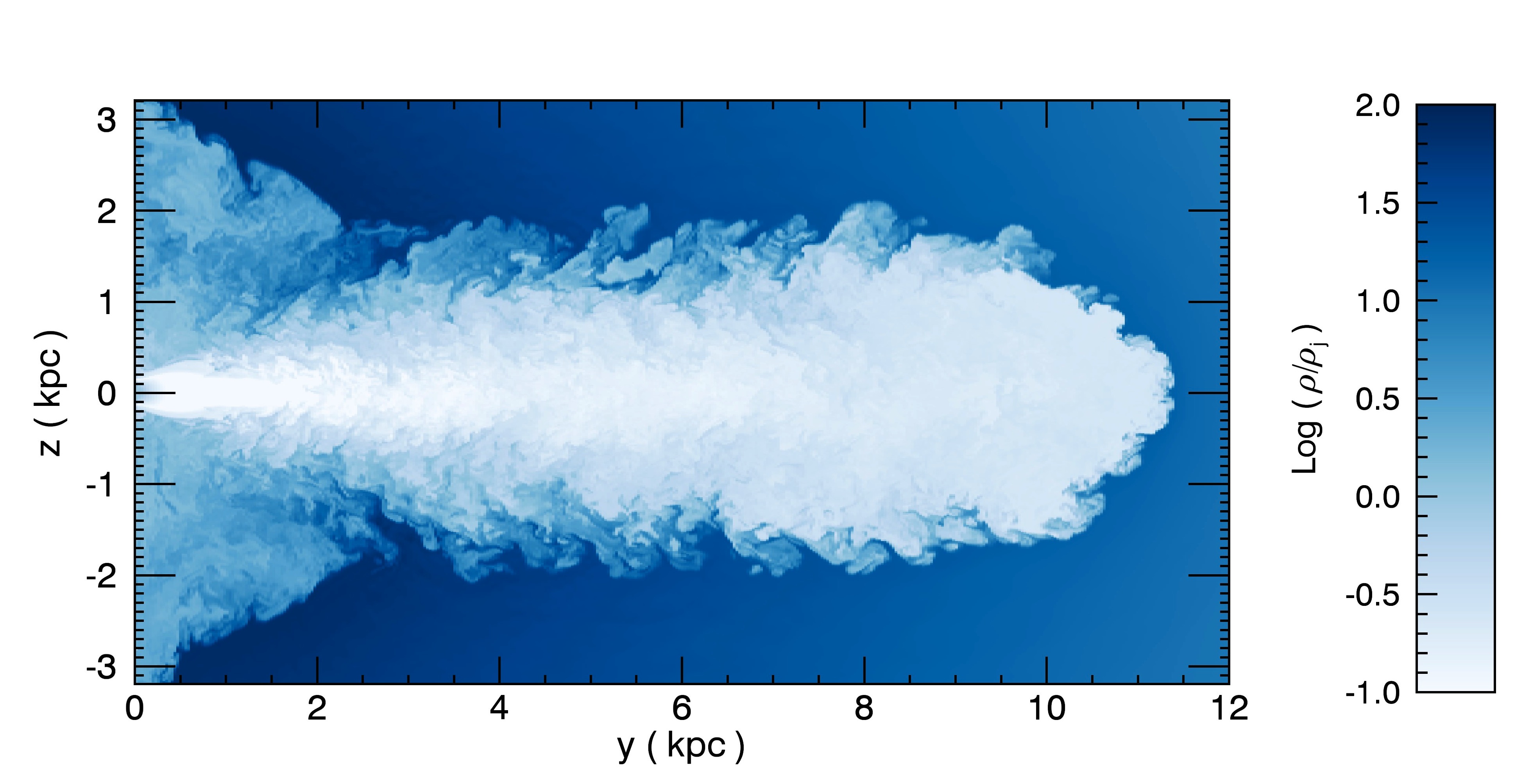} 
\end{center} \caption{ Cut in the $(y,z)$ plane of the logarithmic density distribution for  
case D, overpressured jet, at $t=720$ time units, $1.7 \times 10^7$ yrs.}
\label{fig:D} 
\end{figure*}

In the cases considered so far, we assumed that the jet is initially in
pressure equilibrium with the ambient medium. This choice is motivated by the
fact that an overpressured (underexpanded) jet develops a recollimation shock that leads to
equal pressure between the internal and external gas \citep{belan10}. This probably occurs at radii smaller than those probed by our simulations, within the
initial 100 pc of the jet length, and the jet is therefore already in pressure
equilibrium at injection. It is nonetheless useful to assess  the main
features of the evolution of an overpressured jet. We simulated
this with
$P_{jet}/P_{amb}=10$, as Case D. The general morphology is similar to that of
case A (see Fig. \ref{fig:D}), but, as expected, the jet shows a recollimation
shock at $r<10$ where it has already expanded to twice its initial
radius. This rapid spread causes the jet disruption to occur at smaller
distances, $r\sim 20$, than in case A. Nonetheless, the jet advances
at a very similar speed.

\begin{figure*}[!ht] 
\begin{center} 
\includegraphics[width=2\columnwidth]{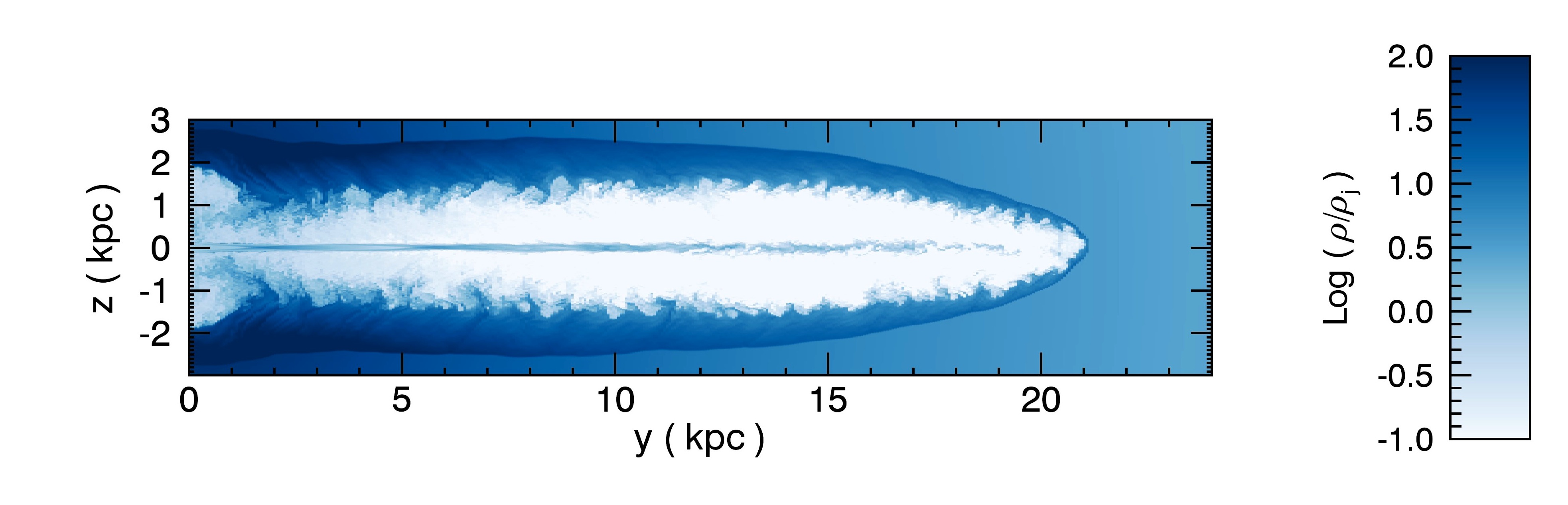} 
\end{center} 
\caption{Cut in the $(y,z)$ plane of the logarithmic density distribution for
  case E, the highest jet power ($M= 40$, $\eta = 10^{-2}$, ${\cal
    L}_{\mathrm kin} = 1.1 \times 10^{45}$), at $t=36$ time units, $3 \times 10^6$ yrs. }
\label{fig:density_E} 
\end{figure*}

\begin{figure*}[!ht] 
\includegraphics[width=1.00\columnwidth]{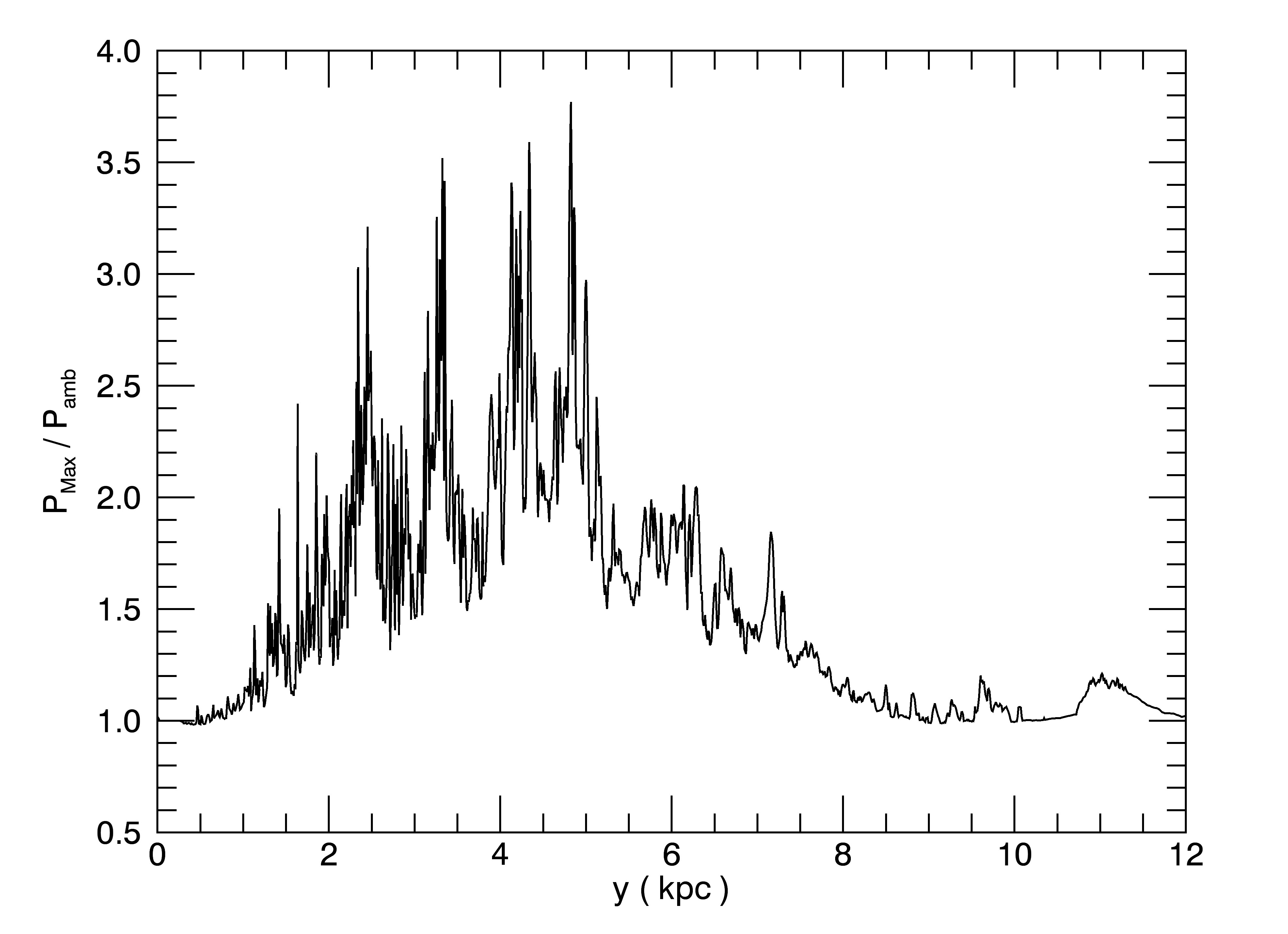} 
\includegraphics[width=1.00\columnwidth]{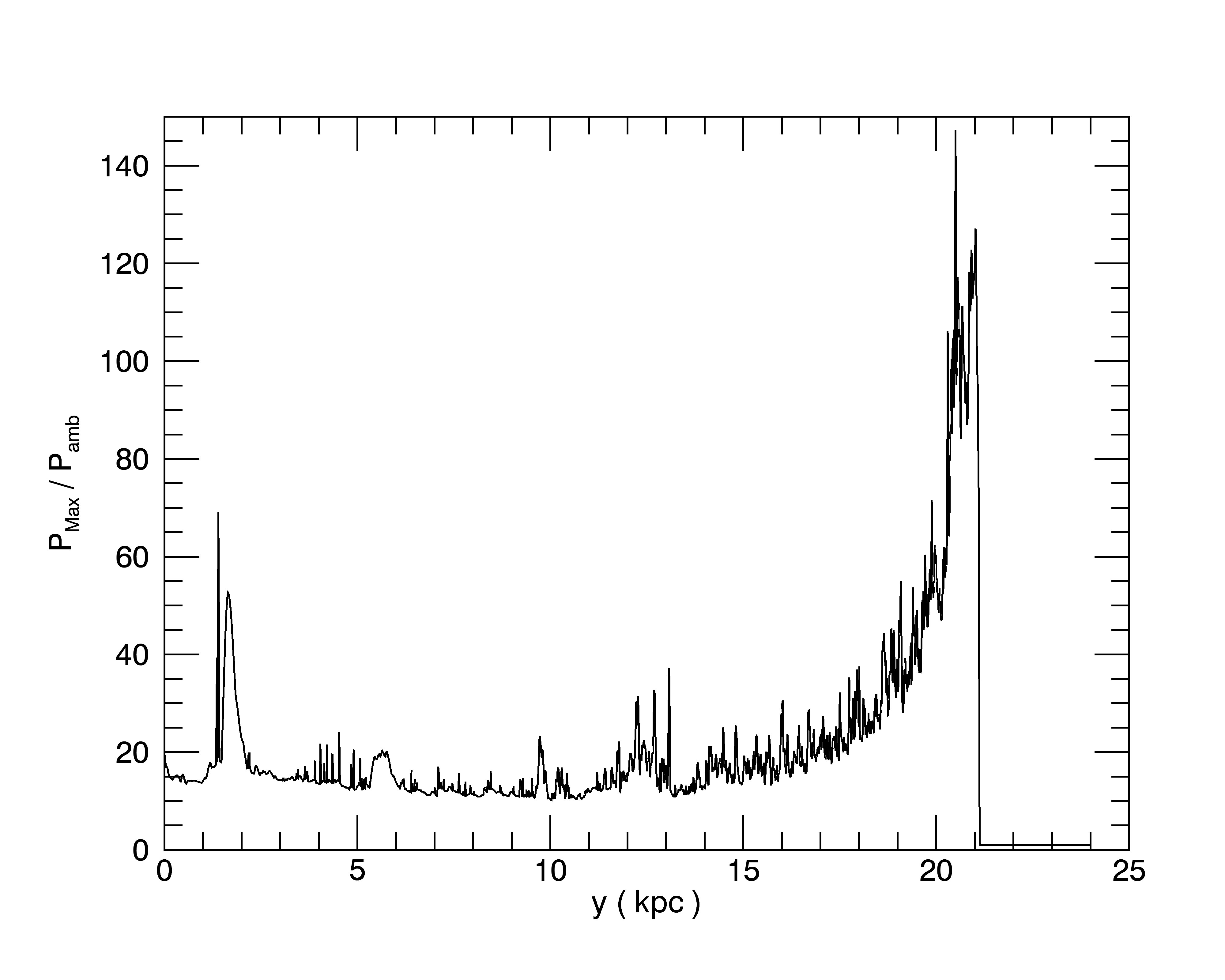}
\caption{Maximum pressure on transverse planes vs $y$ for case B at $t=640$ time units (left
  panel), case E at $t=36$ (right panel).}
\label{fig:pmM4}
\end{figure*}

This investigation is characterized by two main parameters, which
are the density
ratio $\eta$ and the Mach number $M$, as defined in Sect.
\ref{sec:theory}. The cases presented in detail are characterized by the
lowest value of the kinetic power. Exploring the parameter
plane, we reach higher
power and note a transition between FR~I-like to FR~II-like morphologies.
This becomes evident by comparing the density distributions of
Figs. \ref{fig:B} and \ref{fig:density_E} relative to the cases B and E,
respectively. Case E differs from case B only for its Mach number, 40 instead
of 4, but this translates into a 1,000 higher kinetic power. Case E is
characterized by the presence of a shocked region at the jet head, with the
consequent cocoon that shrouds and separates the shocked region by the
external unperturbed medium; these are characteristics typical of FR~II sources.

A more quantitative representation of the differences between these two
simulations is obtained by examining the plots of the maximum pressure as
functions of the longitudinal distance given in Fig. \ref{fig:pmM4}. In the
low-power case B (left panel) the pressure reaches its maximum value at $\sim$
40 jet radii and then it steadily decreases with only a slight increase at the
jet head. In the high-power case E (center panel) the maximum pressure
shows a strong increase toward the jet termination point, in
addition to small-scale peaks (which probably are the locations
of the internal shocks).

\begin{figure*}[!ht] 
\begin{center} 
\includegraphics[width=1.2\columnwidth]{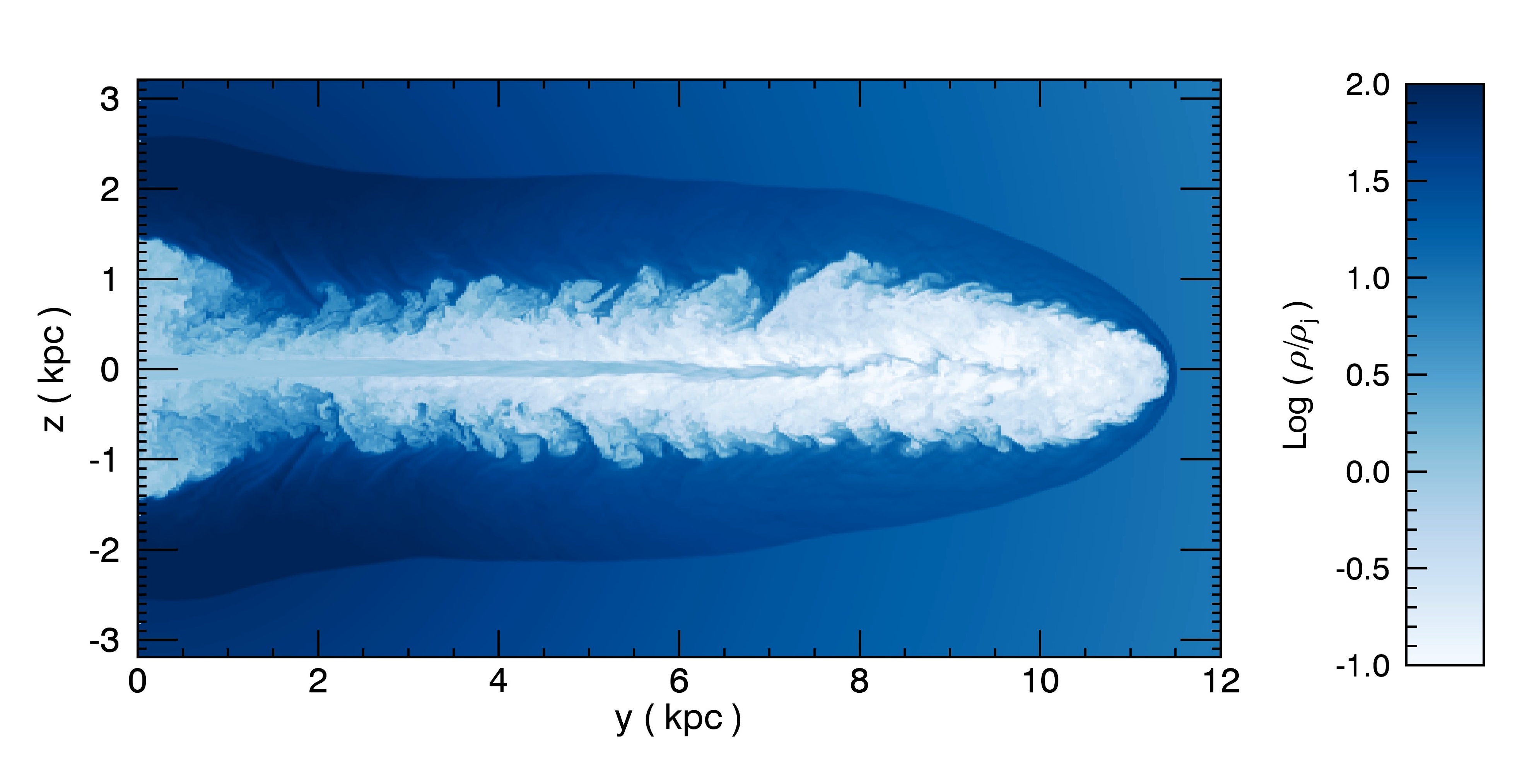}  
\includegraphics[width=0.75\columnwidth]{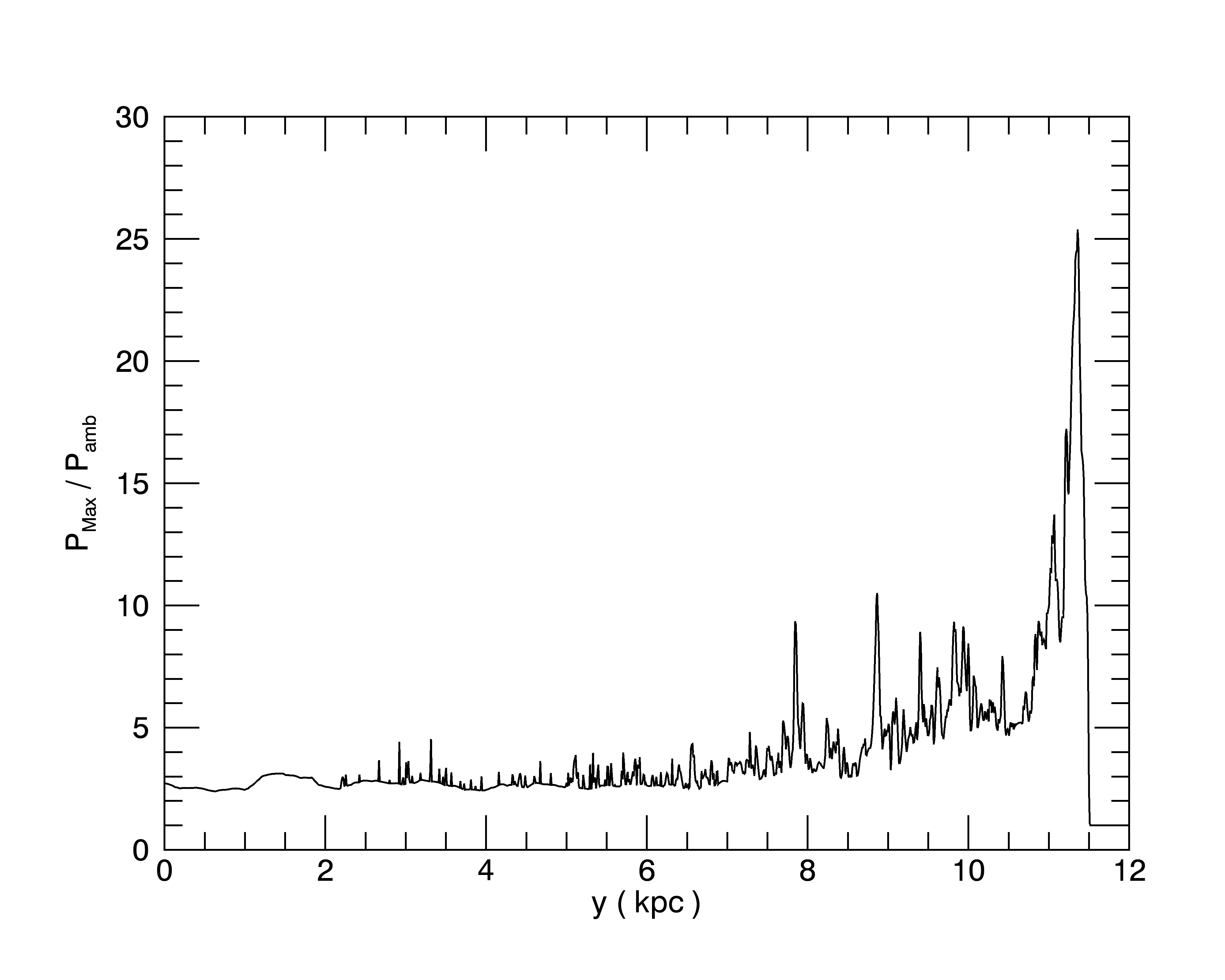} 
\end{center}
\caption{Left: cut in the $(y,z)$ plane of the logarithmic density
  distribution for case F ($M= 10$, $\eta = 10^{-2}$, ${\cal L}_{\mathrm
    kin} = 1.7 \times 10^{43}$) at $t=98$ time units,  $7.5 \times 10^6$ yrs; right: maximum pressure on transverse planes
     along the jet.}
\label{fig:density_H} 
\end{figure*}

At which Mach number does this transition occur, however? Maintaining $\eta$
constant, we performed a simulation with a more moderate Mach number in
case F, $M=10$ (corresponding to ${\cal L}_{\mathrm kin} = 1.7 \times
10^{43}$). Figure \ref{fig:density_H} shows the
cocoon and the shocked region at the jet head, as confirmed by the maximum
pressure behavior in Fig. \ref{fig:pmM4} (right panel), where shocks are still
clearly visible. Thus, at $\eta=0.01$ the transition between FR~I and FR~II
morphologies occurs between $M=4$ and $M=10$.

\begin{figure*}[!ht] 
\begin{center} 
\includegraphics[width=1.2\columnwidth]{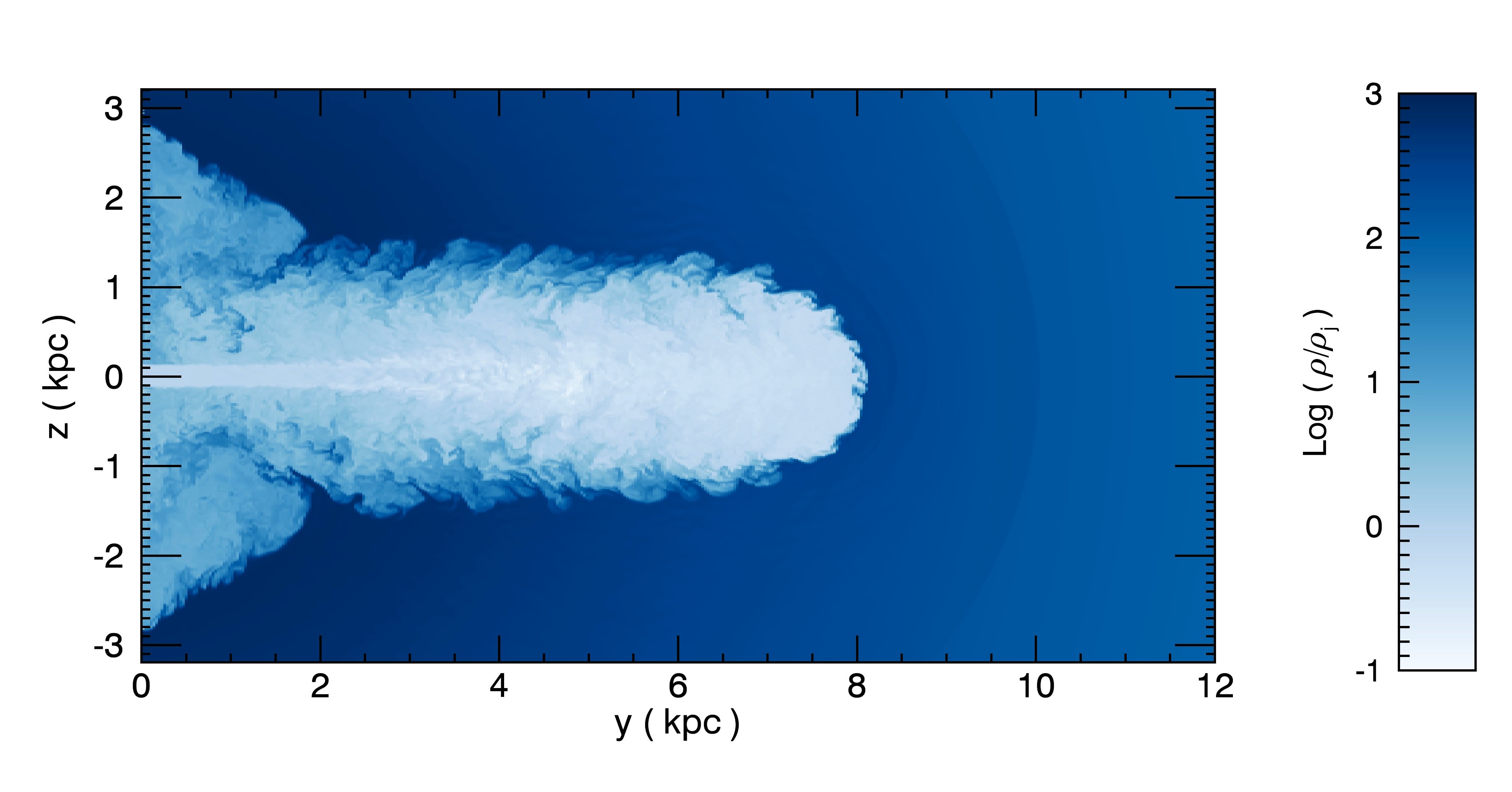} 
\includegraphics[width=0.75\columnwidth]{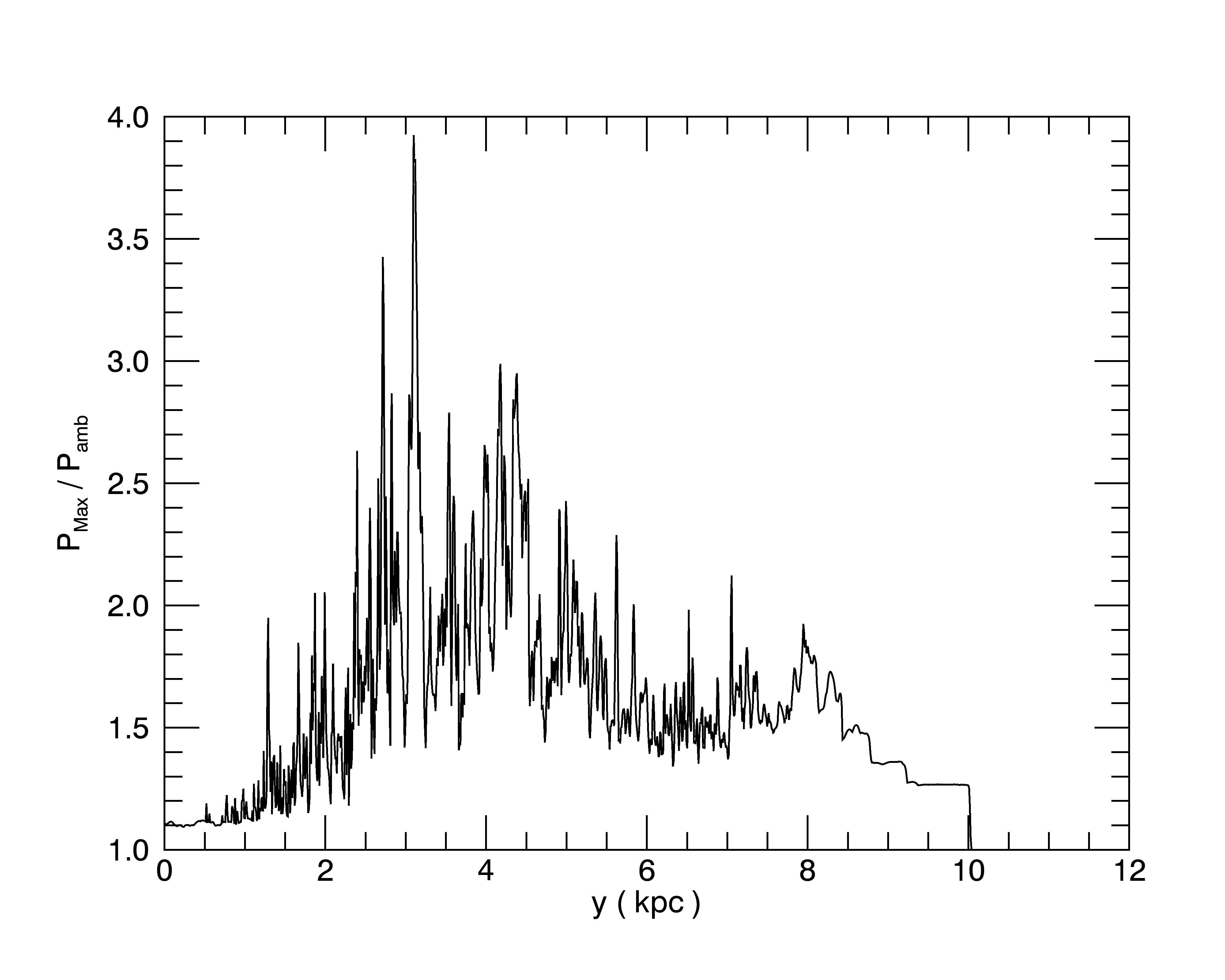} 
\end{center} 
\caption{Left: cut in the $(y,z)$ plane of the logarithmic density
  distribution for case G ($M= 4$, $\eta = 10^{-3}$, ${\cal L}_{\mathrm
    kin} = 3.5 \times 10^{42}$) at $t=1100$ time units, $8.5 \times 10^7$ yrs; right: maximum pressure on transverse planes along
  the jet.}
\label{fig:G} 
\end{figure*}

Finally, we considered case G with the same Mach number ($M=4$) as in the reference
case, but a lower density ($\eta = 10^{-3}$). The jet kinetic power is ${\cal
  L}_{\mathrm kin} = 3.5 \times 10^{42}$, $\sim$3 times higher than case
B. {It is counterintuitive that a lighter jet has a higher jet
  power.  However, from the condition of pressure equilibrium, a lighter jet is also hotter
  and therefore has a higher velocity for the same Mach number.} The resulting density
distribution and pressure profile (see Fig. \ref{fig:G}) indicates that this
case corresponds to a FR~I morphology. 

From our simulations we recover a separation between FR~I and FR~II
morphologies that occurs at a jet power that agrees remarkably
well with the
observations.

\begin{figure*}[!ht] 
\includegraphics[width=0.65\columnwidth]{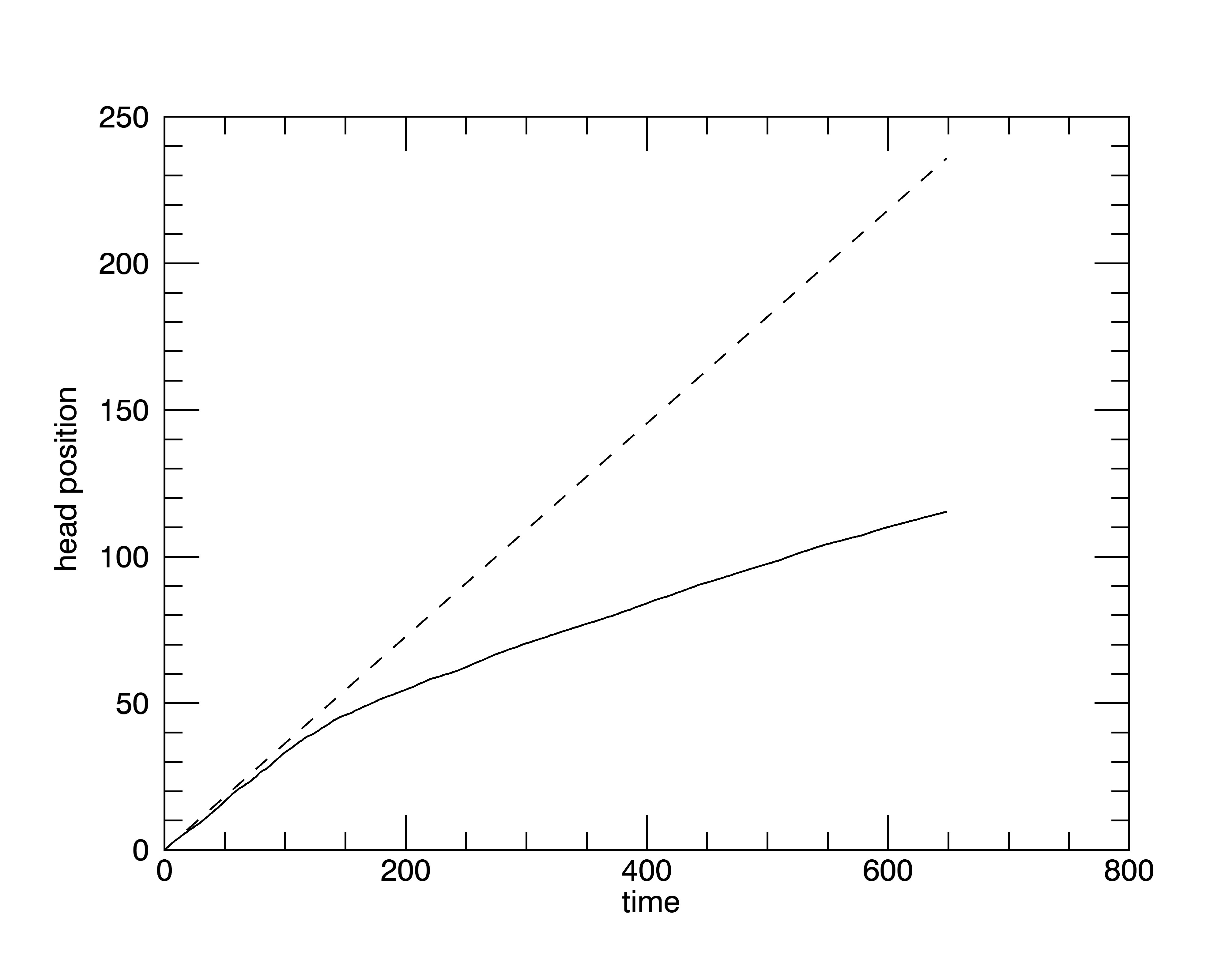}
\includegraphics[width=0.65\columnwidth]{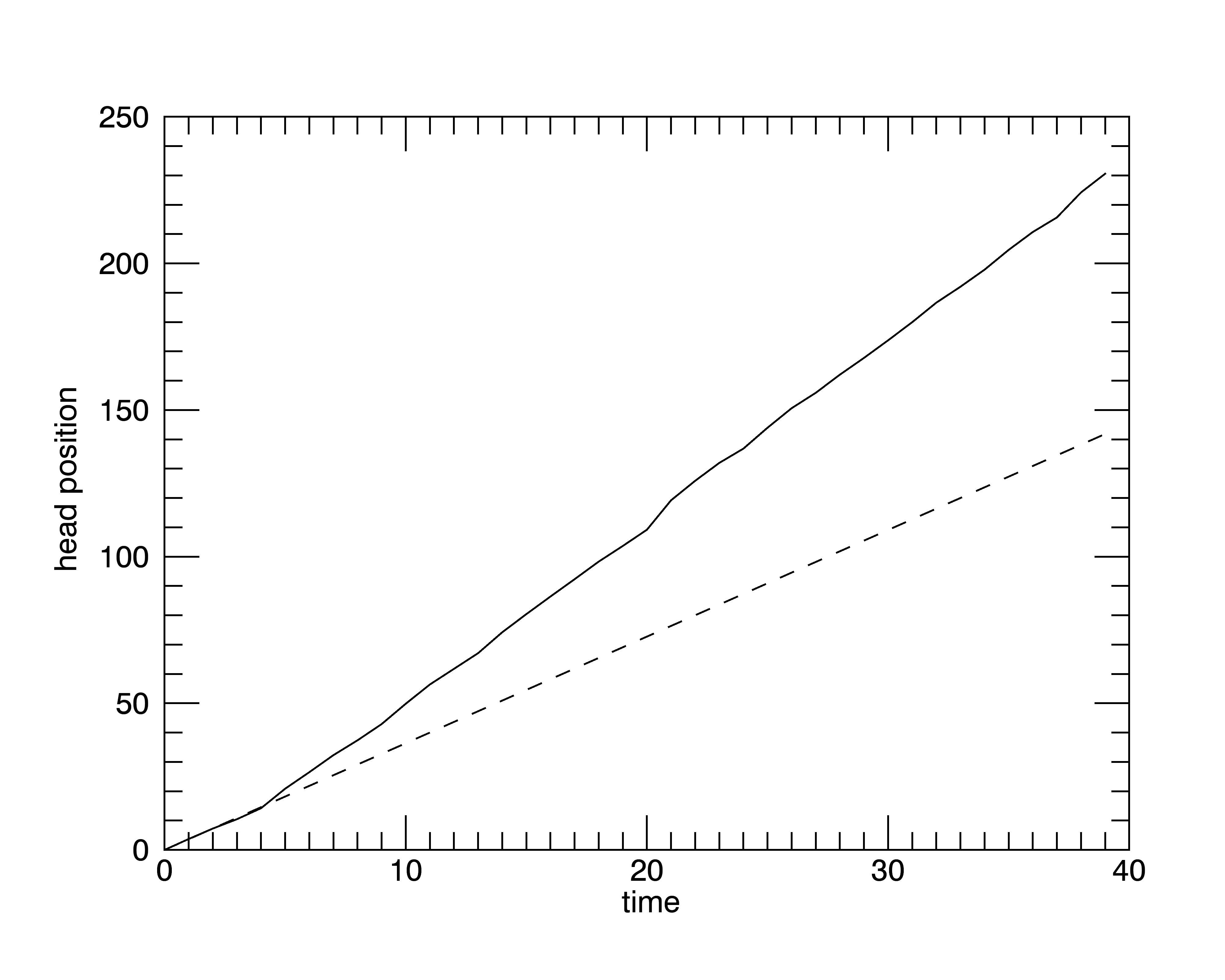}
\includegraphics[width=0.65\columnwidth]{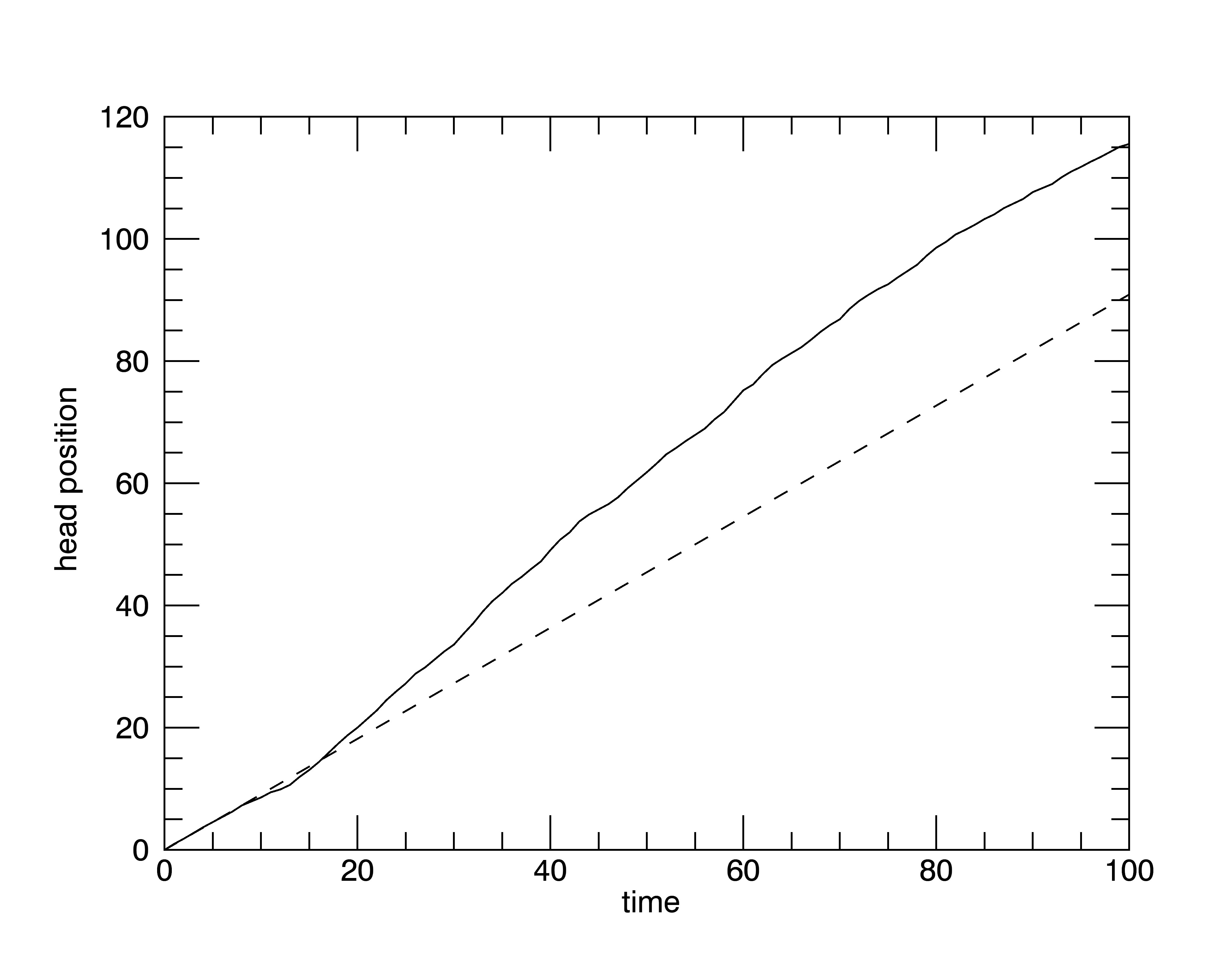} 
\caption{Jet head position as a function of time (solid line) compared with  the theoretical estimate for a uniform medium (dashed line) for cases B, E, and F in the left,
  center, and right panel, respectively.}
\label{fig:av}
\end{figure*}

Another important difference between the cases producing FR~I and FR~II is
the advance speed of the jet. The progress of the jet into the ambient medium
is shown in Fig. \ref{fig:av} for case B (left panel), where the solid
line represents the position of the jet head as a function of time and the
dashed line, given as a reference, is obtained by the longitudinal momentum
conservation at the jet head in a uniform medium \citep{marti94}:

        \begin{equation} y_{\rm head} = \frac{M}{1+1/\sqrt{\eta}} \, t \;. \end{equation}

We note that, notwithstanding the longitudinal decrease of the ambient medium,
which favors an increase in jet propagation speed, the jet slows down
because the Mach disk is disrupted. The jet advance shows a knee
at $t\sim 100$ time units where the velocity drops by a factor $\sim$2. After
the knee the velocity approaches a constant value of $v_{\rm head} \sim 500$km
s$^{-1}$.

Conversely, in cases E and F (Fig. \ref{fig:av}, center and right panel,
respectively) the head progress rate increases when it reaches $r\sim 20$
as a result of the drop in the external density. This is because, unlike what is seen
in case B, the jet remains well collimated. A subtle difference is nonetheless
present as in case F the advance speed appears to decrease at $r\gtrsim 80,$
where indeed the jet starts to loose its coherence, while it remains constant
for the more powerful jet simulated with case E.

\section{Summary and conclusions} 
\label{sec:conclusions}
We have performed three-dimensional numerical simulations of turbulent jets,
which generate morphologies typical of FR~I radio sources. The jet
propagates in a stratified medium that is meant to model the interstellar
intracluster transition.  FR~I radiosources are known to be relativistic at
the parsec scale, therefore a deceleration to sub-relativistic velocities must
occur between this scale and the kiloparsec scale.  In this paper we did not
model the deceleration process, but we assumed that deceleration already
occurred and considered scales where the jet is sub-relativistic.  {  We  did not
consider buoyancy effects, by not taking into account the galaxy gravitational potential, 
but they are expected to be negligible up to the distances reached by the jet head in our
 simulations (of about ten kpc),  while they may become
more relevant at lager distances, as the head further slows down.} In this first
analysis we also neglected the effect of the magnetic field. The parameters
governing the jet evolution are therefore the Mach number $M$ and the initial
jet-to-ambient density ratio $\eta$, which, by constraining the values of the external density and temperature
through observational data, can be
combined to give the jet kinetic power.

The estimated jet kinetic power of the transition between FR~I and FR~II is
$10^{43}$ erg s$^{-1}$ , and we investigated a series of cases below this
threshold.  These low-power cases have $M = 4$ and different values of density
ratio, and they all give rise to turbulent structures typical of FR~I
sources. The jet power, instead of being completely deposited at the
termination through a series of terminal shocks, as in FR~II sources, is
gradually dissipated by the turbulence.  We showed that three-dimensionality
is an essential ingredient for the occurrence of the transition to turbulence.
Two-dimensional simulations with the same parameters lead to FR~II like
behavior with energy dissipation concentrated at the jet
termination. Increasing the Mach number to $40$ and consequently the kinetic
power well above the FR~I - FR~II threshold, we obviously recover well-collimated jets that dump all their energy at the termination shocks. At
intermediate Mach numbers ($M=10$), with a kinetic power around the transition
value, we find characteristics typical of FR~II sources, even though the
energy deposition at the jet termination starts to become more gradual and the
morphology acquires some of the FR~I properties.
 
The simulations presented show that in FR~Is the jet energy is transferred to
the ISM in part inducing, through entrainment, a global low-velocity outflow; the
remaining power is instead dissipated through acoustic waves. The energy released
by active nuclei is thought to play a fundamental role in the evolution of
their host galaxies \citep{fabian12}; in particular, in radio-loud AGN, the
kinetic energy carried by their jets is transferred to the surrounding medium, which
leads to the so-called radio-mode feedback. The FR~I jets, although of
lower power with respect to those of FR~IIs, are extremely important from the
point of view of feedback. This is because the FR~I jets remain confined
within the central regions of the host over longer timescales (and possibly
for their whole lifetime) which exceeds  10$^7$
years for our reference case
B. Furthermore, the entire host is affected by the
radio-mode feedback, while in more powerful radiosources a smaller volume
(immediately surrounding the jets) is involved. Finally, as a
result of the steep radio luminosity function (e.g. \citealt{mauch07}), the less powerful
FR~I radio sources are much more common than the FR~II, and they are then
(potentially) able to affect the general evolution of massive elliptical
galaxies. Clearly, further simulations are required to quantitatively
assess the effects of feedback in FR~Is.

\begin{acknowledgements}
The numerical calculations were performed at CINECA in Bologna, Italy, with the help of an ISCRA grant. PR and GB acknowledge support by PRIN-INAF Grant, year 2014.
\end{acknowledgements}

\bibliographystyle{aa} 
\bibliography{paper_referee}

\end{document}